\documentclass[twocolumn,trackchanges]{aastex63}

\usepackage{amsmath}
\usepackage{units}
\usepackage{upgreek}
\usepackage{xspace}
\usepackage{bm}

\definecolor{orange}{RGB}{253,106,2}
\definecolor{darkgreen}{RGB}{0,100,0}

\graphicspath{{./}{figures/}}

\received{}
\revised{}
\accepted{}

\newcommand{\four}{\unit[450]{\micron}\xspace}
\newcommand{\eight}{\unit[850]{\micron}\xspace}

\newcommand{\msun}{M$_\odot$}

\def\phn{\phantom{0}}       

\submitjournal{ApJ}

\shorttitle{Super-critical filaments in B5}
\shortauthors{Schmiedeke et al.}

\begin{document}

\title{Dissecting the super-critical filaments embedded in the 0.5 pc subsonic region of Barnard 5}

\correspondingauthor{Anika Schmiedeke}
\email{schmiedeke@mpe.mpg.de}

\author[0000-0002-1730-8832]{Anika Schmiedeke}
\affil{Max-Planck-Institut f\"ur extraterrestrische Physik, Gie{\ss}enbachstra{\ss}e 1, 85748 Garching, Germany}

\author[0000-0002-3972-1978]{Jaime E. Pineda}
\affiliation{Max-Planck-Institut f\"ur extraterrestrische Physik, Gie{\ss}enbachstra{\ss}e 1, 85748 Garching, German y}

\author[0000-0003-1481-7911]{Paola Caselli}
\affiliation{Max-Planck-Institut f\"ur extraterrestrische Physik, Gie{\ss}enbachstra{\ss}e 1, 85748 Garching, Germany}

\author[0000-0001-5653-7817]{Héctor G. Arce}
\affiliation{Department of Astronomy, Yale University, P.O. Box 208101, New Haven, CT 06520-8101, USA}

\author[0000-0001-8509-1818]{Gary A. Fuller}
\affiliation{Jodrell Bank Centre for Astrophysics, Department of Physics and Astronomy, The University of Manchester, Oxford Road, Manchester M13 9PL, UK}
\affiliation{Intituto de Astrof\'isica de Andalucia (CSIC), Glorieta de al Astronomia s/n E-18008, Granada, Spain}

\author[0000-0003-1312-0477]{Alyssa A. Goodmann}
\affiliation{Center for Astrophysics, Harvard \& Smithsonian, 60 Garden Street, Cambride, MA 02138, USA}

\author[0000-0002-7026-8163]{María José Maureira}
\affiliation{Max-Planck-Institut f\"ur extraterrestrische Physik, Gie{\ss}enbachstra{\ss}e 1, 85748 Garching, Germany}

\author[0000-0003-1252-9916]{Stella S. R. Offner}
\affiliation{Department of Astronomy, The University of Texas at Austin, Austin, TX 78712, USA}

\author[0000-0003-3172-6763]{Dominique Segura-Cox}
\affiliation{Max-Planck-Institut f\"ur extraterrestrische Physik, Gie{\ss}enbachstra{\ss}e 1, 85748 Garching, Germany}

\author[0000-0002-0368-9160]{Daniel Seifried}
\affiliation{Universität zu Köln, I. Physikalisches Institut, Zülpicher Str. 77, 50937 Köln, Germany}

%
%
%

\begin{abstract}

We characterize in detail the two \unit[$\sim$0.3]{pc} long filamentary structures found within the subsonic region of Barnard 5. We use combined GBT and VLA observations of the molecular lines NH$_3$(1,1) and (2,2) at a resolution of \unit[1800]{au}, as well as JCMT continuum observations at \unit[850 and 450]{\micron} at a resolution of \unit[4400]{au} and \unit[3000]{au}, respectively. We find that both filaments are highly super-critical with a mean mass per unit length, $M/L$, of \unit[$\sim$80]{\msun pc$^{-1}$}, after background subtraction, with local increases reaching values of \unit[$\sim$150]{\msun pc$^{-1}$}. This would require a magnetic field strength of \unit[$\sim$500]{$\mu$G} to be stable against radial collapse.
We extract equidistant cuts perpendicular to the spine of the filament and fit a modified Plummer profile as well as a Gaussian to each of the cuts. The filament widths (deconvolved FWHM) range between \unit[6500-7000]{au} (\unit[$\sim$0.03]{pc}) along the filaments. This equals $\sim$2.0 times the radius of the flat inner region. We find an anti-correlation between the central density and this flattening radius, suggestive of contraction. Further, we also find a strong correlation between the power-law exponent at large radii and the flattening radius. We note that the measurements of these three parameters fall in a plane and derive their empirical relation. Our high-resolution observations provide direct constraints of the distribution of the dense gas within super-critical filaments showing pre- and protostellar activity.

\end{abstract}


%
%

\section{Introduction} \label{sec:intro}

Dense cores are the places where stars form \citep[see reviews by][]{diFrancesco2007,WardThompson2007,Bergin2007}. They present non-thermal subsonic velocity dispersions \citep{Fuller1992,Goodman1998,Caselli2002} and represent the end of the turbulent cascade \citep{Larson1981}. 

One of the most popular molecular tracers to study the dense gas within and around dense cores is ammonia, NH$_3$. This is because the NH$_3$(1,1) transition traces material with densities starting at a few \unit[$10^3$]{cm$^{-3}$}. Also, due to its hyperfine structure, it is a useful determinant of optical depth and kinematic properties. In addition the inversion transitions NH$_3$(1,1) and NH$_3$(2,2) usually can be observed simultaneously, as they are only separated by \unit[28.1]{MHz}. The observation of these two transitions provide measurements of temperatures and column density \citep{Friesen2017}.

By mapping dense cores in NH$_3$(1,1), it was found that they show an almost constant level of non--thermal motions, within a certain ``coherence'' zone (Goodman et al. 1998). The term ``coherent core'' describes the dense gas where non-thermal motions are roughly constant, and typically smaller than the thermal motions, independent of scale \citep[see also][]{Caselli2002}.

Wide field dust continuum observations with the {\em Herschel Space Observatory} have revealed that filaments are commonplace throughout molecular clouds, with lengths ranging from \unit[0.5]{pc} up to several tens of pc. It was also established that they host most of the dense cores, with bound ones located predominantly in filaments with transcritical to super-critical masses per unit length \citep[e.g.][]{Andre2014,Arzoumanian2018, Arzoumanian2019}. From these dust continuum observations, it is argued that the width of those filaments are close to 0.1 pc \citep{Arzoumanian2011, Arzoumanian2019}.

Studies using molecular line observations have also identified filamentary structures. Since molecular lines provide kinematic information, the filamentary structures are sometimes also analyzed taking into account their coherence in velocity. Identified filaments using molecular lines usually have smaller length 
than those filaments seen by {\em Herschel}. A single several pc long filament identified with {\em Herschel} can show several \citep[in a few cases intertwined;][]{Hacar2013, Henshaw2013} velocity-coherent filamentary structures when studied through molecular lines \citep{FernandezLopez2014,Hacar2017,Suri2019,Chen2020}. \citet{Suri2019} find that the filament widths in the Orion A molecular cloud varies between \unit[0.02 and 0.3]{pc} using C$^{18}$O(1-0), while filaments as narrow as \unit[0.01]{pc} have been found in the OMC-1 region using NH$_3$(1,1) observations \citep{Monsch2018}. Regarding their stability, the velocity-coherent structures or `fibers' in \citet{Hacar2013,Hacar2017}, were shown to have masses per unit length close to or below the stability value. 
Similar smaller scales filaments were also identified within the coherent, subsonic region of Barnard 5 \citep[hereafter B5;][]{Pineda2011a}. At least one of the narrower filaments found in B5 displays an averaged radial profile that is different than those derived for the {\em Herschel} filaments.

Although efforts have been made to study the fragmentation of the filaments from the theoretical \citep[e.g.][]{Fiege2000, Hanawa2017, Hanawa2019} and the numerical side \citep[e.g.][]{Smith2014, Smith2016, Tomisaka2014, Kirk2015, Seifried2015, Clarke2019, Heigl2020}, the observational efforts have been focused on the fragments themselves \citep[e.g.][]{Pineda2015, Kainulainen2016} and an observationally driven understanding of filament fragmentation has yet to be established.
The analysis of observations is usually performed on the average density profile. Hence it is unclear if and how the density profile and width of a filament that is undergoing fragmentation changes along the spine of the filament (i.e. its main axis). A robust observational determination of these properties along the spine of a filament would provide a strong constrain to theories of filament formation and evolution.

In this work, we focus on a region located in the Perseus star-forming region at a distance of \unit[(302 $\pm$ 21)]{pc} from the Sun \citep{Zucker2018}. This region, called B5, has nearly constant subsonic non-thermal velocity dispersion, covering an area of \unit[$\sim$ 0.4]{pc} $\times$ \unit[0.6]{pc} and hosts at least one young stellar object (YSO), B5-IRS1 \citep{Fuller1991}. High angular resolution (\unit[6]{\arcsec}, \unit[1800]{au}) NH$_3$ observation reveals filamentary substructures within the subsonic region \citep{Pineda2011a}. Embedded in these filaments are, in addition to the already mentioned YSO, three gravitationally bound dense gas condensations \citep{Pineda2015}, forming a wide-separation quadruple system. This quadruple system appears to be the result of fragmentation of the dense gas filaments. Here, we use these high angular resolution, high density tracer observations of NH$_3$ to study in detail the density profile of the filamentary structure embedded in the coherent zone of B5.

The paper is structured as follows. In Sec.~\ref{sec:data} we present the observational data used in this paper. In Sec.~\ref{sec:results} we report the results of the analysis of the filamentary structure. These results are discussed in Sec.~\ref{sec:discussion}. We conclude the paper in Sec.~\ref{sec:summary}.

%
%
\section{Observations}\label{sec:data}

We use observational data from three different telescopes: Continuum maps at \unit[450 and 850]{\micron} obtained with the \emph{James-Clerk-Maxwell Telescope (JCMT)}, and combined observations of two metastable ammonia transitions, NH$_3$(1,1) and (2,2), obtained with the \emph{Robert C. Byrd Green Bank Telescope (GBT)} and the \emph{Very Large Array (VLA)}. These observational data have already been published \citep{Pineda2015}. For clarity we briefly list the important details of these observations.

\subsection{James-Clerk-Maxwell Telescope (JCMT)}\label{ssec:jcmt}

Observations at \four and \eight of the B5 cloud have been performed using the Submillimetre Common-Use Bolometer Array 2 \citep[SCUBA-2][]{Holland2013} at the JCMT (project code: M13BU14) during grade 1 weather. The \four and \eight observations were carried out simultaneously on 2013 August 16 and 23, and 2013 September 3. The iterative map-making technique is used with the command \verb+makemap+ \citep{Chapin2013} of the Starlink software suite\footnote{\url{http://starlink.eao.hawaii.edu/starlink}}, with a pixel size of \unit[0.5]{\arcsec} to match the NH$_3$(1,1) VLA map (see Sec. \ref{ssec:gbt_vla}). Details of the data reduction can be found in \citet{Pineda2015}. The maps have a spatial resolution of \unit[9.8]{\arcsec} (\unit[3000]{au}) at \four and \unit[14.6]{\arcsec} (\unit[4400]{au}) at \eight. The noise level in the emission free regions is \unit[0.23]{mJy pixel$^{-1}$} at \four and \unit[0.026]{mJy pixel$^{-1}$} at \eight.

\subsection{Robert C. Byrd Green Bank Telescope (GBT) and Very Large Array (VLA)} \label{ssec:gbt_vla}
GBT observations of the B5 region have been carried out between 2009 December 23 and 2010 March 21 (project number 08C-088). Two \unit[12.5]{MHz} windows have been centered on NH$_3$(1,1) and NH$_3$(2,2), and observed in frequency switching mode. The spectral resolution of the data is \unit[0.04]{km s$^{-1}$}. Details on the data processing can be found in \citet{Pineda2010}.

The single-dish GBT data have been combined with high resolution interferometric data obtained with the VLA. The VLA observations have been carried out in the D-array configuration on 2011 October 16-17 and in the DnC-array configuration on 2012 January 13-14 (project number 11B-101). The WIDAR correlator was configured such that 2 basebands with \unit[4]{MHz} bandwidth were centered on NH$_3$(1,1)f and NH$_3$(2,2), with a spectral resolution of \unit[0.049]{km s$^{-1}$}. Detailed information on the data processing can be found in \citet{Pineda2015}. The spatial resolution of the final map is \unit[6.0]{\arcsec} (\unit[1800]{au}).



\section{Results} \label{sec:results}

\subsection{Morphology of the subsonic region in B5}

\begin{figure*}[t]
    \includegraphics[width=0.98\textwidth]{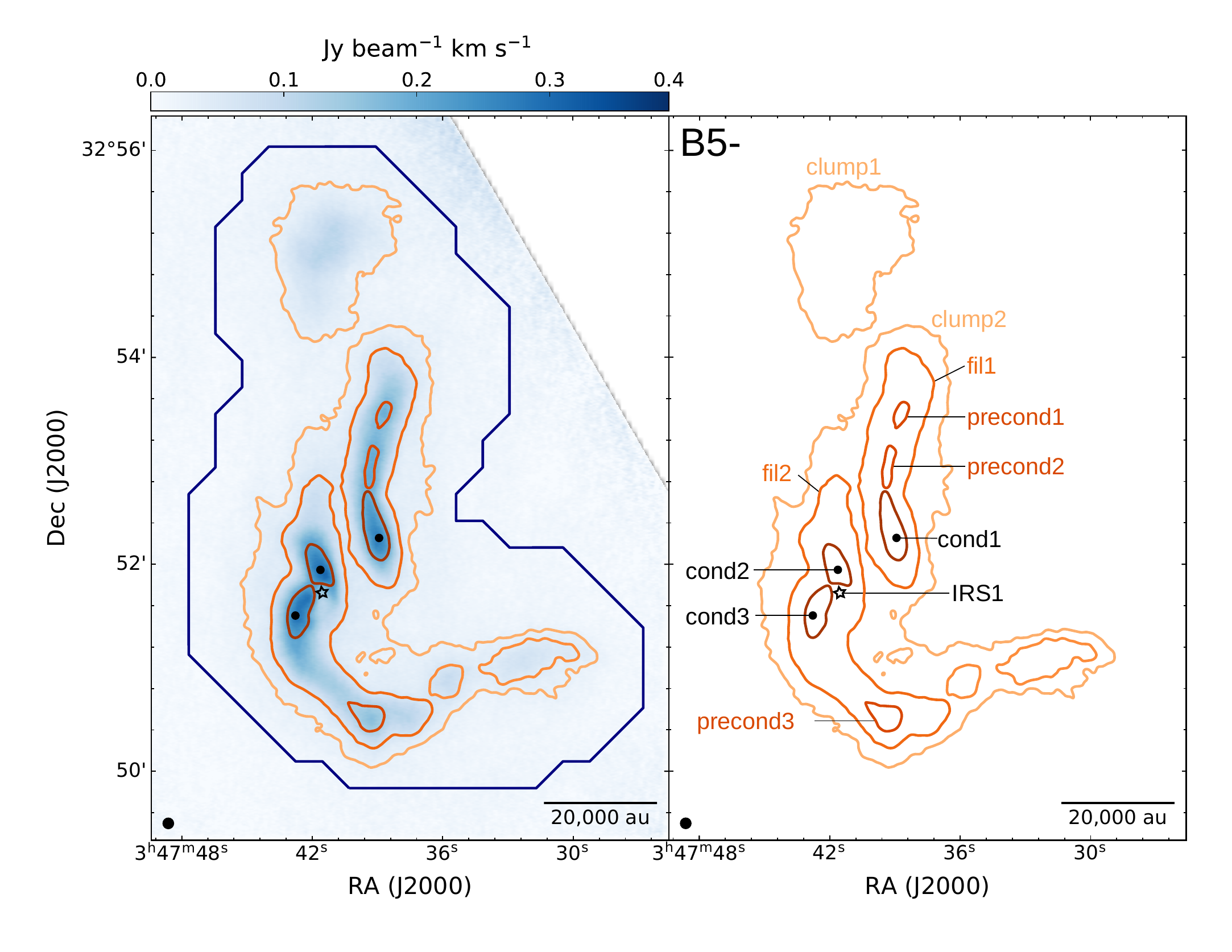}
    \caption{\emph{Left:} Integrated intensity map of VLA and GBT combined NH$_3$(1,1) is shown in the background. The navy contour indicates the extent of the coherent zone \citep{Pineda2010}.
    \emph{Right:} Boundaries and nomenclature for the regions used in this work. The yellow, orange and red contours indicate the structures identified using dendrograms. The black star and the black circles mark the locations of the protostar B5-IRS1 and the gas condensations, respectively \citep{Pineda2015}. The beam and scalebar are shown in the bottom left and right corner, respectively.} 
    \label{fig:structure}
\end{figure*}

The substructure of the B5 coherent region is more easily identified in the high-contrast NH$_3$ integrated intensity image than in the {\em Herschel} images, since NH$_3$ does not trace the more extended structure seen in the {\em Herschel} dust continuum emission due to NH$_3$ either being not present or having very low abundance.

We characterise the morphology of the coherent region employing \verb+astrodendro+ \citep{Rosolowsky2008} on the NH$_3$(1,1) integrated intensity map. By using a clipping method \verb+astrodendro+ assigns emission above a certain threshold (contour level) as associated with an object.  We provide the minimum value to be considered in the map (\unit[0.03]{Jy beam$^{-1}$ km s$^{-1}$}, i.e. 10 times the rms), the height threshold value that determines if a leaf will be a single entity or not (\unit[0.01]{Jy beam$^{-1}$ km s$^{-1}$}), and a minimum number of pixels for a leaf to be considered a single entity (\unit[250]{pixels}).
Fig.~\ref{fig:structure} shows the identified substructure with different contours.

Two independent clumps are identified: B5-clump1 and B5-clump2. While B5-clump1 appears smooth and without signs of active star-formation (quiescent), B5-clump2 breaks up further and shows clear signs of fragmentation. We identify two independent filamentary structures: B5-fil1 and B5-fil2. Along the spines of both filamentary structures we identify three leafs, representing three gas condensations B5-cond1, B5-cond2, and B5-cond3. This finding is in agreement with \citet{Pineda2015}, in which the three condensations along with the protostellar source B5-IRS1 were found to be bound and on their way to form a wide separation quadruple system. Furthermore, here we identify lower level over-densities along both filament spines, which might provide the seeds for future fragmentation (pre-condensations). The peak intensities of these pre-condensations are about 25 \% lower than the faintest condensation.


\begin{figure*}[t]
    \centering
    \includegraphics[width=0.98\textwidth]{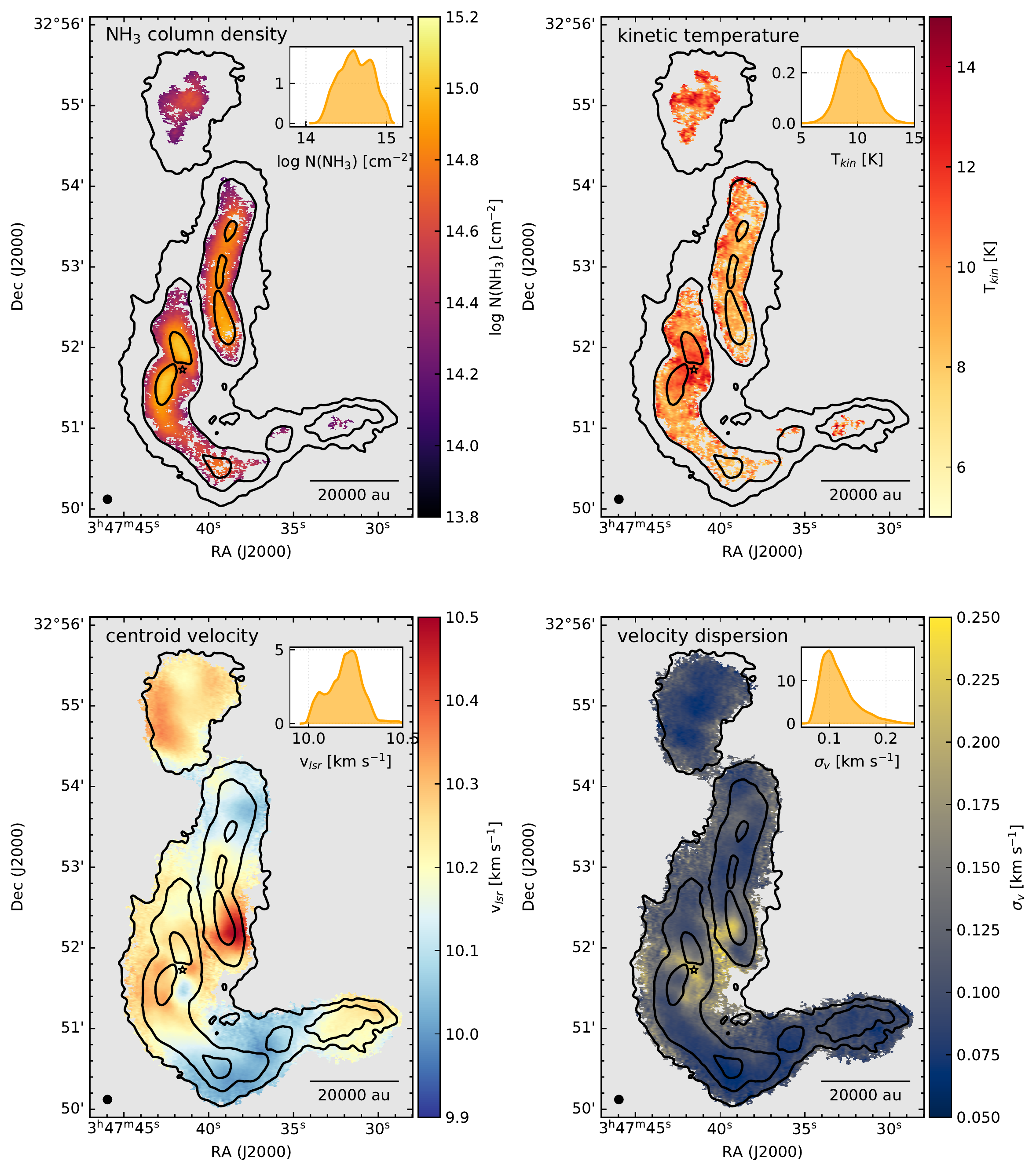}
    \caption{Fit results from simultaneous fitting of NH$_3$(1,1) and (2,2) using pySpecKit. \emph{Top left:} NH$_3$ column density in logarithmic scale. \emph{Top right:} Kinetic temperature, T$_\text{kin}$. \emph{Bottom left:} Centroid velocity $v_\text{lsr}$. \emph{Bottom right:} Velocity dispersion, $\sigma_v$. The black contours are the same as in Fig.~\ref{fig:structure}. The inset in the top right corner of each map shows the distribution of the respective parameter. The beam and scale bar are shown in the bottom left and right corner of each map, respectively. The black star marks the location of the protostar B5-IRS1. Details of the line fitting procedure are given in Appendix~\ref{app:line_fitting}.} 
    \label{fig:B5_nh3_fit_propertyMaps}
\end{figure*}

\begin{deluxetable*}{c|cccc}[]
    \tablecaption{Averaged properties of the substructure of the coherent region of B5, shown in Fig.~\ref{fig:structure}. \label{tab:czStructure}}
    \tablehead{ %
        \colhead{ID} & 
        \colhead{$<$log(N(NH$_3$))$>$} &
        \colhead{$<T_\text{kin}>$} &
        \colhead{$<v_\text{lsr}>$} &
        \colhead{$<\sigma_\text{nt}>$}\\
        \colhead{[B5-]}& 
        \colhead{[cm$^{-2}$]} &
        \colhead{[K]} &
        \colhead{[km/s]} &
        \colhead{[km/s]}
    }
    \startdata
    \hline
    clump1      & 14.42 $\pm$ 0.12 &     10.7 $\pm$ 1.3 &  10.26 $\pm$ 0.05 & 0.06 $\pm$ 0.02\\
    clump2      & 14.65 $\pm$ 0.20 &      9.7 $\pm$ 1.3 &  10.18 $\pm$ 0.09 & 0.09 $\pm$ 0.04 \\\hline
    fil1        & 14.66 $\pm$ 0.18 & \phn 9.2 $\pm$ 1.2 &  10.22 $\pm$ 0.11 & 0.10 $\pm$ 0.04 \\
    fil2        & 14.65 $\pm$ 0.19 & \phn 9.9 $\pm$ 1.3 &  10.19 $\pm$ 0.09 & 0.09 $\pm$ 0.04 \\\hline
    precond1    & 14.82 $\pm$ 0.03 & \phn 8.4 $\pm$ 1.0 &  10.13 $\pm$ 0.01 & 0.08 $\pm$ 0.01 \\
    precond2    & 14.85 $\pm$ 0.02 & \phn 8.5 $\pm$ 0.9 &  10.22 $\pm$ 0.01 & 0.08 $\pm$ 0.01 \\
    precond3    & 14.70 $\pm$ 0.05 & \phn 8.3 $\pm$ 0.6 &  10.05 $\pm$ 0.03 & 0.04 $\pm$ 0.01 \\\hline
    cond1       & 14.87 $\pm$ 0.05 & \phn 8.6 $\pm$ 0.7 &  10.36 $\pm$ 0.08 & 0.14 $\pm$ 0.04 \\
    cond2       & 14.94 $\pm$ 0.06 &     10.8 $\pm$ 0.7 &  10.24 $\pm$ 0.03 & 0.12 $\pm$ 0.03 \\
    cond3       & 14.94 $\pm$ 0.06 &     10.6 $\pm$ 0.9 &  10.28 $\pm$ 0.03 & 0.11 $\pm$ 0.03 \\\hline
    \enddata
\end{deluxetable*}

\subsubsection{General properties of the substructures}

We simultaneously fit the NH$_3$(1,1) and NH$_3$(2,2) data\-cubes using \verb+pySpecKit+ \citep{Ginsburg2011} to obtain information on the NH$_3$ column density, the kinetic temperature and the velocity field. Details of this fitting procedure are given in Appendix~\ref{app:line_fitting}. The resulting property maps are shown in Fig.~\ref{fig:B5_nh3_fit_propertyMaps}.

For each substructure, we list the mean properties obtained from the \verb+pySpecKit+ fitting in Tab.~\ref{tab:czStructure}. This includes the NH$_3$ column density, kinetic temperature, line-of-sight velocity and non-thermal velocity dispersion. The non-thermal velocity dispersion $\sigma_\text{nt}$  is calculated from the fitted velocity dispersion $\sigma_\text{v}$ as 

\begin{equation}\label{eq:sigma_v}
  \sigma_\text{nt} = \sqrt{\sigma_\text{v}^2 - \sigma_\text{th}^2},
\end{equation}
where $\sigma_\text{th}$ is the thermal velocity dispersion.

We notice that the filaments have a lower mean kinetic temperature compared to the clumps. It is also interesting to note that B5-cond1 which is embedded in B5-fil1 has a mean kinetic temperature of \unit[8.5]{K}, while the other two condensations B5-cond2 and B5-cond3 have higher mean kinetic temperatures of \unit[10.7]{K}. These two condensations are embedded in B5-fil2, relatively close to the YSO B5-IRS1 and likely affected by the stellar feedback. The line-of-sight velocities are comparable for all structures, suggestive of shallow velocity gradients along the line-of-sight within the coherent zone of B5. The non-thermal velocity dispersion is sub-sonic as expected. The velocity dispersion map shows a subtle increase at the position of the condensations and the centroid velocity map shows variations where the additional clumpy substructure appears.

\subsection{Filamentary structure}\label{ssec:filaments}

\subsubsection{Filament length and profiles perpendicular to filament spines}\label{ssec:fitting}

\begin{figure}[t]
    \includegraphics[width=0.48\textwidth]{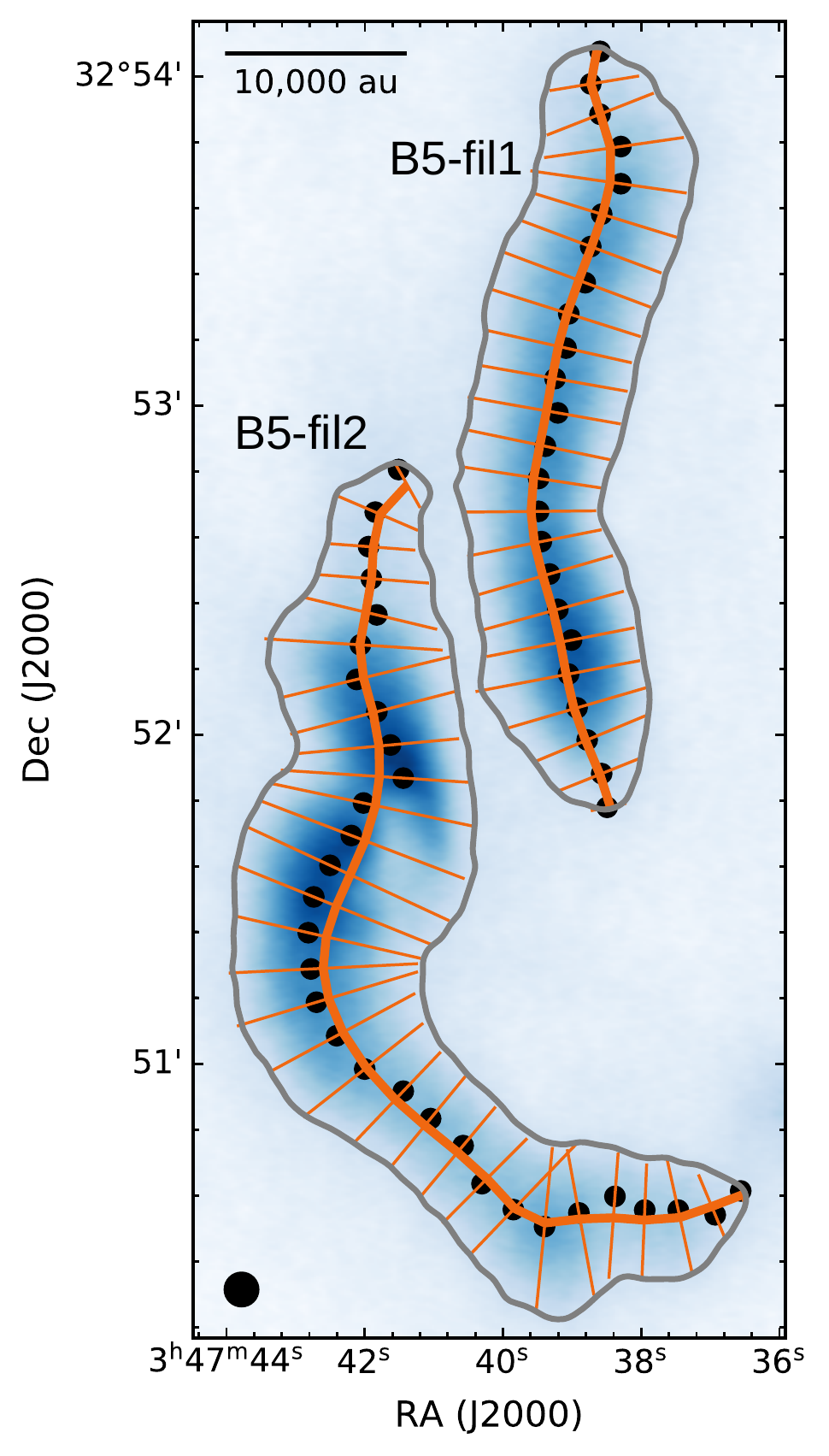}
    \caption{Ammonia integrated intensity map of the masked filaments. The spines of the filaments are marked by the solid thick orange lines. The perpendicular cuts are marked in solid thin orange lines. The peak pixel of each intensity cut is indicated by a black circle. The beam is indicated by the black circle in the bottom left corner.} 
    \label{fig:filamentSpines}
\end{figure}

We employ the python-based package \verb+radfil+ \citep{Zucker2018a} to further characterize both filamentary structures. Based on the provided NH$_3$(1,1) intensity map and a mask, radfil employs the package \verb+fil_finder+  \citep{Koch2015} to define the spine of the filaments. Following the spines of the filament, we determine filament lengths of \unit[0.24]{pc} and \unit[0.31]{pc} for B5-fil1 and B5-fil2, respectively. Going along the filament spine, we extract equidistant cuts perpendicular to the spine using a sampling frequency of 12 pixels. This yields roughly a one beam separation between individual cuts (\unit[6]{\arcsec} beam, \unit[0.5]{\arcsec} pixelsize). The resulting profiles are shifted such that the center coincides with the peak intensity. The spine and the perpendicular cuts are shown in Fig. \ref{fig:filamentSpines}.

%

To investigate the filament properties, we follow the approach presented by \citet{Arzoumanian2011} and adopt the idealized model of a cylindrical filament with a radial density $n(r)$ and column density $\Sigma(R)$ profile as

\begin{multline}\label{eq:plummer}
    n(r) = \frac{n_0}{\left\lbrack 1+ \left(r/R_\text{flat}\right)^2\right\rbrack^{p/2}} + n_\text{bkg} \\  \longrightarrow 
    \Sigma(R) = A_p \frac{\Sigma_0}{\left\lbrack 1 + \left(R/R_\text{flat}\right)^2\right\rbrack^\frac{p-1}{2}} + \Sigma_\text{bkg},
\end{multline}
where 
\begin{equation}\label{eq:central_density}
    \Sigma_0 = n_0 R_\text{flat}. 
\end{equation}
Here $r$ is the cylindrical radius from the spine of the filament and $R$ is the projected radius, $n_0$ is the central density of the cylinder, $R_\text{flat}$ is the radius of the flat inner section of the cylinder, $n_\text{bkg}$ and $\Sigma_{bkg}$ are the (constant) background density and surface density, respectively, and $A_p$ is a finite constant factor that depends on the density profile power-law index $p$ and the filament inclination angle (which we assume to be equal to zero for simplicity). For $p=2$ the filament width is $W \sim 3 R_\text{flat}$ \citep{Arzoumanian2011}. But more commonly, the filament width is defined as the full-width-at-half-maximum, FWHM, of a Gaussian fit to inner part of the cuts \citep[e.g.][]{Arzoumanian2011,Panopoulou2017}. Hence in addition to fitting a Plummer function, we also fit a Gaussian to the innermost (\unit[8000]{au}) part of the cut, taking into account the constant background emission. 

\begin{equation}\label{eq:gauss}
    g(R) = \frac{A_\text{G}}{\sigma_\text{G}\sqrt{2\pi}}\exp{\left(-\frac{(R - \mu_\text{G})^2}{2\sigma_\text{G}^2}\right)} + g_\text{bkg},
\end{equation}
where $A_\text{G}$ is the amplitude, $\sigma_\text{G}$ is the variance, and $\mu_\text{G}$ is the expected value. From this the FWHM is calculated as $\text{FWHM} = 2\sigma_\text{G}\sqrt{2 \ln 2}$. The beam is taken into account by deconvolving the FWHM, following \citet{Koenyves2015}, i.e. the deconvolved FWHM, $FWHM_\text{d} = \sqrt{(FWHM^2 - HPBW^2)}$, where HPBW is the half-power beamwidth in au. For our observations, the HPBW is \unit[6]{\arcsec}, corresponding to \unit[$\sim$ 1800]{au} at the distance of B5. For the Plummer fits, we convolve the Plummer-like function with the \unit[6]{\arcsec} Gaussian beam prior to fitting.

\begin{figure}[t]
    \centering
    \includegraphics[width=0.48\textwidth]{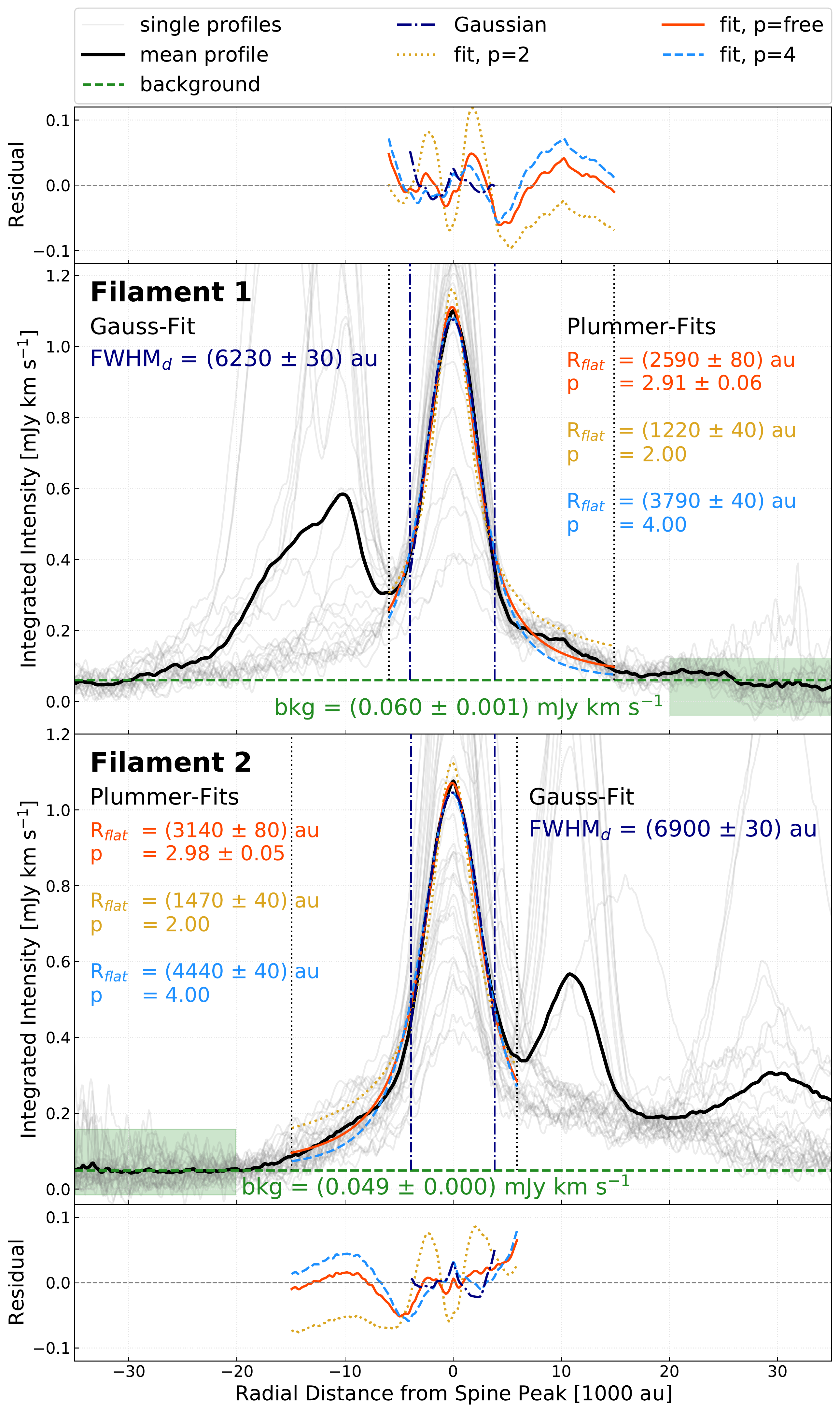}
    \caption{Results from fitting the averaged profiles (solid black) with a Gaussian profile (navy dashed-dotted line, which mostly overlaps with the averaged profile) and a Plummer profile, for which (i) the exponent is fixed to 2 (gold dotted line), (ii) the exponent is fixed to 4 (light blue dashed line) and (iii) the exponent is kept as a free parameter (orange solid line). \emph{Top:} B5-fil1. \emph{Bottom:} B5-fil2. The light gray lines are the individual profile cuts. The secondary peak in the averaged profile (on the left for B5-fil1 and on the right for B5-fil2) is the contamination of the profile cut from the respective other filament. This has been excluded from the fitting. The fitting range is indicated by the vertical navy dashed-dotted lines for the Gaussian fit and by the vertical black dotted lines for the Plummer fits.
    The horizontal dashed green line indicates the background emission, obtained by fitting the outer edge of the profile cuts; only the un-contaminated side of the averaged profile is used for the background determination (marked with the light green box).} 
    \label{fig:filamentProfiles}
\end{figure}

To investigate the global properties of the filaments, we first fit the average profiles of each filament. In Fig.~\ref{fig:filamentProfiles}, we show the averaged profiles in solid black. The individual profiles are superimposed in light gray. When fitting the profiles, we restrict the fitting range to avoid contamination from peak structures from the other filament. The background emission is determined by fitting a constant to the flat outer edges of the profile cuts. We fit three different versions of Eq. \ref{eq:plummer}, (i) the exponent is fixed to 2, (ii) the exponent is kept as a free parameter, and (iii) the exponent is fixed to 4. The latter is the \citet{Ostriker1964} solution for an isothermal filament in hydrostatic equilibrium. The best-fit of all three versions are shown in Fig. \ref{fig:filamentProfiles} in gold, orange, and blue respectively. 

We list the fitting results in Tab.~\ref{tab:widths_comparison} in the appendix (data set `NH$_3$, \unit[6]{\arcsec}'). We use the Akaike Information Criterion (AIC) to evaluate the goodness of the fits, where the model with the lowest AIC value is the preferred one. For both filaments, the lowest AIC value is determined for the fit with a Plummer function where the exponent has been kept as a free parameter.

We then fit each individual perpendicular cut with a Gaussian and a Plummer-like profile where the exponent is kept as a fitting parameter. From this fitting, we obtain the variation of the filament parameters (mass, central density, width and exponent) as a function of location along the filament spines. These results are presented in detail in the following subsections.

%

\subsubsection{Filament mass}\label{ssec:mass}

We follow two independent strategies to determine the mass of the filaments: (a) Scaling the NH$_3$(1,1) integrated emission using the JCMT \four map and (b) converting the NH$_3$ column density to mass. The detailed description of the procedures and comparison of both conversions is given in Appendix~\ref{app:convFactors}. In short, for method (a), we convert the JCMT \four background-corrected\footnote{We determine the background as the pixel within the structure with the lowest value.} flux density in B5-cond1 to gas mass assuming a gas-to-dust ratio of 100 \citep{Hildebrand1983}, a distance of \unit[302 $\pm$ 21]{pc} \citep{Zucker2018}, a dust temperature of \unit[9 $\pm$ 1]{K} (deduced from the kinetic temperature map and assuming that the dust temperature is coupled to the gas temperature), and optically thin dust with a dust absorption coefficient of $\kappa_{450\mu m}$ = \unit[6.4 $\pm$ 0.81]{cm$^2$g$^{-1}$} \citep{Ossenkopf1994}. For the same condensation, B5-cond1, we measure the background-corrected$^2$ flux density in the integrated NH$_3$(1,1) map. This yields an ammonia-to-mass conversion factor of \unit[(1.5 $\pm$ 0.7)]{\msun Jy$^{-1}$}. For method (b) we convert the ammonia column density derived from the \verb+pySpecKit+ fitting assuming an abundance of ammonia with respect to H$_2$ of 10$^{-8.5}$ \citep{Friesen2017}. The resulting mass maps agree within a factor of two with each other (method (b) yields a factor of two higher mass compared to method (a)). The ratio of both maps is smooth and does not show any strong gradients. All mass-dependent calculations in this work will be performed on the basis of the mass map derived with the dust-scaling method (a), since it is a conservative estimate.

For the entire filaments B5-fil1 and B5-fil2, we measure background corrected flux densities of \unit[6.29]{Jy} and \unit[8.97]{Jy}, respectively in the NH$_3$(1,1) integrated intensity map. We determine the background as the pixel within the filament with the lowest flux density (\unit[$\sim$0.4]{mJy} for both filaments). This is very likely overestimating the real background and hence it results in a very conservative mass determination. We obtain total background corrected masses of $M_\text{B5-fil1}$ = \unit[9.4]{\msun} and $M_\text{B5-fil2}$ = \unit[13.4]{\msun}.

\subsubsection{Filament mass per unit length}

We determine the mass per unit length, $M/L$, to be \unit[39.2]{\msun pc$^{-1}$} for B5-fil1 and \unit[43.2]{\msun pc$^{-1}$} for B5-fil2. We note that this is a conservative estimate due to the aggressive background determination.

A more realistic approach encompasses determining the local variation of filament mass. Here, for each cut perpendicular to the filament spine, the background is determined by fitting a constant to the outer edges of the cuts. The determined background values are \unit[0.05 -- 0.06]{mJy}. We subtract this background when calculating the mass enclosed in the filament cut. To obtain the mass per unit length, we divide the mass contained in the individual cut by the length of the cut in the direction of the filament spine. 
In panel (b) of Fig.~\ref{fig:B5_fil_properties} we show the variation of filament background corrected mass per unit length along the filament spine. The mean value is \unit[74.2]{\msun pc$^{-1}$} for B5-fil1 and \unit[83.3]{\msun pc$^{-1}$} for B5-fil2, which we also list in Tab. \ref{tab:filProps}. Due to the $\sim$1-beam separation between the cuts and the filaments being rather straight, the overlap of individual cuts and hence the duplication when determining the mass, is minimal (see Fig.~\ref{fig:filamentSpines}).

The critical value $(M/L)_\text{crit}$ for an isothermal cylinder of gas in hydrostatic equilibrium is \citep{Ostriker1964}:
\begin{equation}\label{eq:isoMline}
    (M/L)_\text{crit} = \frac{2c_s^2}{G} \sim 16.6 \left(\frac{T_\text{gas}}{\unit[10]{K}}\right)\,\text{M}_\odot\,\text{pc}^{-1},
\end{equation}
where $c_s$ is the sound speed of the gas, $G$ is the gravitational constant, and $T_\text{gas}$ is the gas temperature. Both filaments exceed this critical value on average by a factor of 3-4. Locally this can even increase by factors of 5. Both filaments are supercritical, which means that their thermal pressure is insufficient to support them against gravitational collapse. They should collapse further unless there are means providing additional support (e.g. magnetic fields). This is discussed in Sec.~\ref{ssec:bfield}.

\subsubsection{Central density}\label{ssec:central_density}

Similar to the determination of the flux-to-mass conversion factor $\xi$ outlined in Sec. \ref{ssec:mass}, we calculate a flux-to-column-density factor $\Xi$ as

\begin{equation}
    \Xi = \frac{\mathcal{G}}{\kappa_\nu\, \mu_{H_2}\, m_H\, \Omega\, B_\nu(T_d)},
\end{equation}
where $\mathcal{G}$ is the gas-to-dust ratio \citep[$\mathcal{G} = 100.$;][]{Hildebrand1983}, $\kappa_\nu$ is the dust absorption coefficient at the frequency $\nu$, $\mu_{H_2}$ is the molecular weight per hydrogen molecule \citep[$\mu_{H_2}$=2.8;][]{Kauffmann2008}, $m_H$ is the mass of the hydrogen atom, $\Omega$ is the area (in our case of B5-cond1) and $B_\nu(T_d)$ is the Planck function evaluated at the dust temperature $T_d$.
We determine a \four flux-to-column-density factor of $\Xi = (4.4 \pm 1.9) 10^{25} \text{Jy}^{-1} \text{cm}^{-2}$. From the flux measured for B5-cond1 in both the JCMT \four and the NH$_3$(1,1) map, we determine an ammonia-to-column-density conversion factor of \unit[$\eta$ = (1.25 $\pm$ 0.53) 10$^{26}$]{Jy$^{-1}$cm$^{-2}$}.

We fit the observed intensity profile of the cuts perpendicular to the filament spine with a modified Plummer function (similar to Eq.~\ref{eq:plummer}, see Sec.~\ref{ssec:fitting}), where the fitting parameter $A$ is given as:
\begin{equation}\label{eq:convCentralDensity}
    A = \frac{n_0 \, A_p \, R_\text{flat}}{\eta}.
\end{equation}
This allows us to determine the central density $n_0$.
From the fit to the averaged profile of the filament cuts, we determine a central density of \unit[(1.7 $\pm$ 0.7)~10$^6$]{cm$^{-3}$} for B5-fil1 and \unit[(1.4 $\pm$ 0.6)~10$^6$]{cm$^{-3}$} for B5-fil2. 
In the panel (c) of Fig.~\ref{fig:B5_fil_properties}, we show the variation of central density along the filament spines for both filaments, determined from the individual fits to the filament profiles. The mean of the central density derived from the individual fits is \unit[1.8$\times$ 10$^6$]{cm$^{-3}$} for B5-fil1 and \unit[1.5 $\times$~10$^6$]{cm$^{-3}$} for B5-fil2, just slightly higher compared to the value determined from fitting the averaged profile (see also Tab.~\ref{tab:filProps}). As expected, we see an increase in central density towards the location of all three condensations. We note that the profiles of the variation of central density along the filament spines visually appear to be similar when one flips one profile and aligns the condensations. This is further evaluated in Sec.~\ref{ssec:sims}.
The central density, on the order of \unit[10$^6$]{cm$^{-3}$}, is comparable to central densities found in pre-stellar cores like L1544 \citep{Crapsi2005}.

\subsubsection{Filament width and exponent}\label{ssec:width_exp}

%

From the Gaussian fit of the averaged profiles, we obtain a deconvolved FWHM$_\text{d}$ of \unit[(6200 $\pm$ 30)]{au} for B5-fil1 and \unit[(6900 $\pm$ 30)]{au} for B5-fil2, which corresponds to $\sim 2R_\text{flat}$ for the Plummer fits, where the exponent $p$ has been kept as a free parameter. Given their lengths, we determine aspect ratios, i.e. length over FWHM$_\text{d}$, of 17:1 and 10:1 for B5-fil1 and B5-fil2, respectively. 

\begin{figure*}[t]
    \centering
    \includegraphics[height=0.94\textheight]{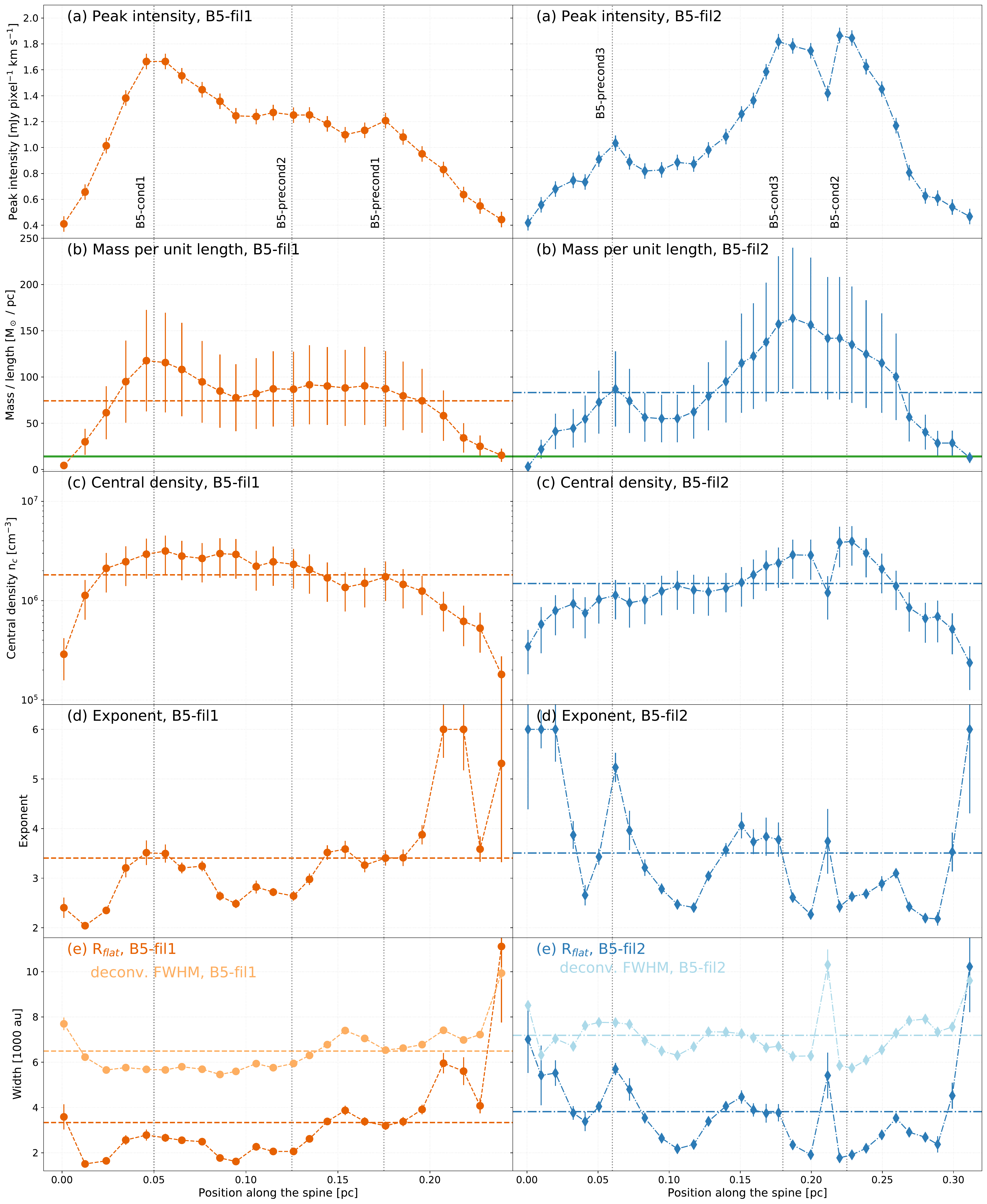}
    \caption{Properties along the spines of the filaments. \emph{Left:} B5-fil1, marked with orange circles and a dashed line. \emph{Right:} B5-fil2, marked with blue diamonds and a dash-dotted line . From top to bottom: (a) spine peak intensity, (b) mass per unit length; the solid green line indicates the critical limit determined by \citet{Ostriker1964} for a \unit[10]{K} cylinder with subsonic turbulence. (c) central density, (d) exponent $p$ from Eq.~\ref{eq:plummer}, (e) $R_\text{flat}$ and deconvoled FWHM in lighter colors. The dashed and dash-dotted horizontal lines mark the average values of the spine properties. The location of the condensations and the pre-condensations are marked by the vertical gray dotted line and labelled in the top panel. The position 0.0 starts at the South of each spine.} 
    \label{fig:B5_fil_properties}
\end{figure*}

We then apply the same Gaussian fitting to each individual cut along the filament spines as described in Sec.~\ref{ssec:fitting}. Panel (e) in Fig.~\ref{fig:B5_fil_properties} shows the variation of $R_\text{flat}$ and FWHM$_\text{d}$ along the spines of both filament. The mean of FWHM$_\text{d}$ is \unit[6500]{au} for B5-fil1 and \unit[7200]{au} for B5-fil2 (see also Tab. \ref{tab:filProps}). Both values agree well with the results from fitting the average profile. The deviation could be caused by the higher uncertainty in the fits due to the lower signal-to-noise towards the tips of the filaments. We discuss the determined filament widths further in Sec.~\ref{ssec:widths}.

The mean of $R_\text{flat}$ is \unit[3300]{au} for B5-fil1 and \unit[3800]{au} for B5-fil2. Hence the mean of  $R_\text{flat}$ increases by about 20 -- 30 \%, when fitting the individual profiles as opposed to fitting the averaged profiles. This increase is likely linked to the increase in the exponent. Fitting the averaged profile with a Plummer profile where the exponent has been kept as a free parameter yields exponent values of $p\approx3$. When fitting the individual profiles, we determine a mean value of $p\approx3.5$.
A value of $p=4$ represents a special case of an isothermal filament in hydrostatic equilibrium  \citep{Ostriker1964}. Lower values of $p$, $1.5<p<2.5$, are typically found in molecular cloud filaments like e.g. IC5146 \citep{Arzoumanian2011} using dust continuum emission maps. \citet{Monsch2018} on the other hand report the detection of a narrow filament, $R_\text{flat}=\unit[0.013]{pc}$ (i.e \unit[2680]{au}), with a steep exponent, $p=5.1$, in the Orion A OMC1 region using the dense gas tracer NH$_3$.
Along the spine, we notice a trend of decreasing $R_\text{flat}$ towards the condensations, i.e. towards increasing $n_0$. Similarily, we notice an increase in the exponent when $R_\text{flat}$ increases. These (anti-)correlations are further discussed in Sec.~\ref{ssec:evolution}.

\subsubsection{Kinematics}

\begin{figure*}[t]
    \centering
    \includegraphics[width=0.98\textwidth]{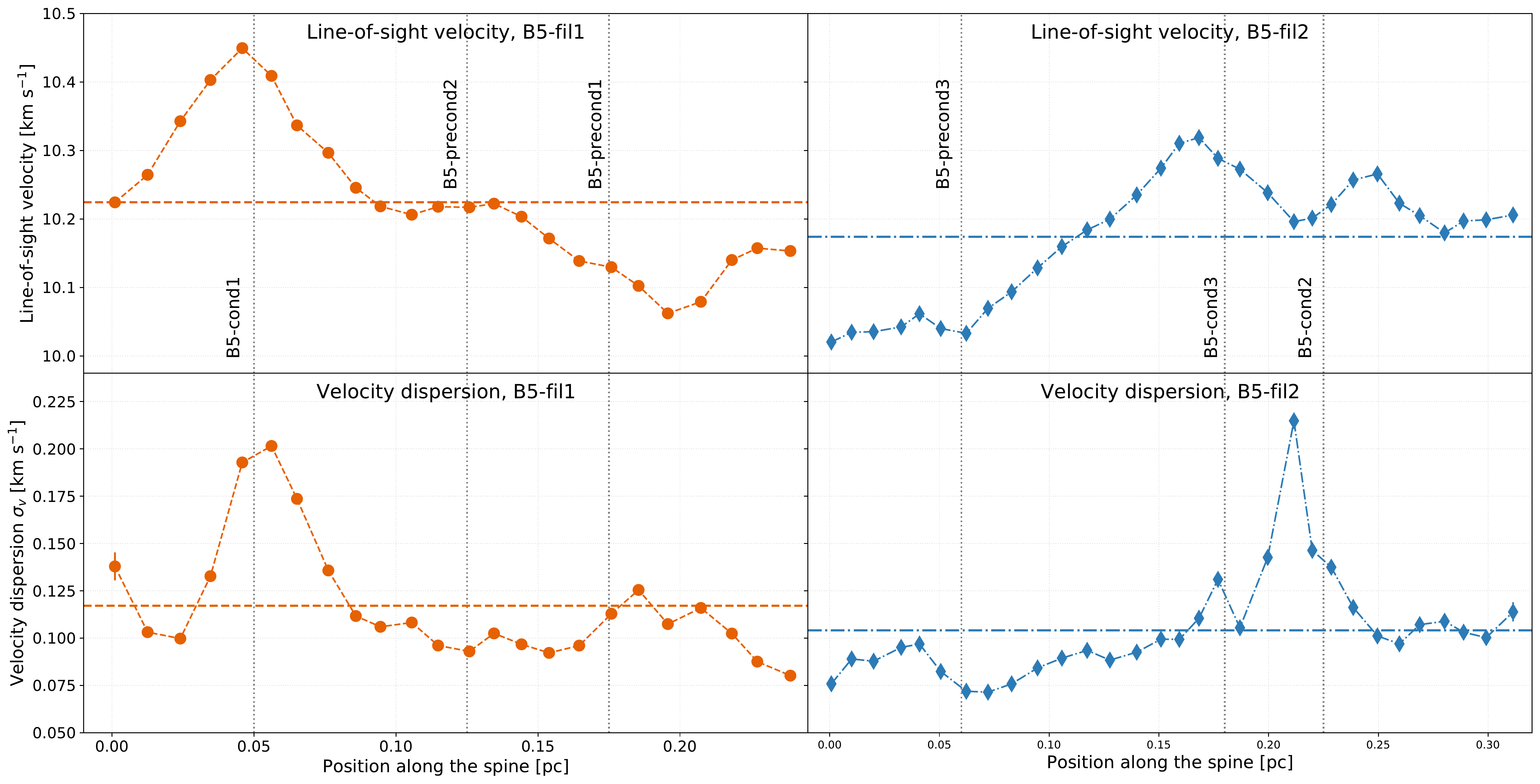}
    \caption{Kinematic properties along the spines of the filaments. \emph{Left:} B5-fil1, marked with orange circles and a dashed line. \emph{Right:} B5-fil2, marked with blue diamonds and a dash-dotted line. \emph{Top:} Line-of-sight velocity $v_\text{lsr}$. \emph{Bottom:} Velocity dispersion $\sigma_\text{v}$.
    The uncertainties for the majority of the datapoints are located within the markers. The dashed and dash-dotted horizontal lines mark the average values of the spine properties. The vertical gray dotted lines mark the location of the three condensations as well as the pre-condensations. The position 0.0 starts at the South of the spine.}
    \label{fig:vlsr}
\end{figure*}

In Fig.~\ref{fig:vlsr} we show the line-of-sight velocity, $v_\text{lsr}$ and the velocity dispersion $\sigma_\text{v}$ along both filament spines. There is a mild gradient in correspondance with the condensations, suggestive of inflowing material. This has also been suggested by \citet{Hacar2011}, where they show the velocity oscillations along filamentary structures in Taurus. It is noticeable that B5-cond1 falls at a peak, while B5-cond2 and B5-cond3 do not. This could be related to the protostar B5-IRS1 affecting the line-of-sight velocity structure in B5-fil2. The velocity dispersion $\sigma_\text{v}$, including both thermal and non-thermal motions, shows an increase in B5-fil1 towards B5-cond1 and in B5-fil2 towards the location in between the other two condensations. This might be caused by the protostellar object B5-IRS1, which is located nearby.

%

%

\begin{deluxetable}{l|c c}[tbp]
    \tablecaption{Derived filament properties. \label{tab:filProps}}
    \tablehead{
    \colhead{Parameter} &
    \colhead{B5-fil1} &
    \colhead{B5-fil2}
    }
    \startdata
    \textbf{Flux density $F$ [Jy]}                              &           &  \\
        \hspace{.5cm} $F_\text{fil}$                            & 11.95     & 17.50 \\
        \hspace{.5cm} $F_\text{fil,bkg-sub}$                    & \phn 6.29 & \phn 8.97 \\ \hline
    \textbf{Mass $M$ [\msun]}                                   &           &  \\
        \hspace{.5cm} $M_\text{fil}$                            & 17.9      & 26.1 \\
        \hspace{.5cm} $M_\text{fil,bkg-sub}$                    & \phn 9.4  & 13.4 \\ \hline
    \textbf{Length $L$ [pc]}                                    & 0.24      & 0.31 \\ \hline
    \textbf{$M/L$ [\msun pc$^{-1}$]}                            &           &  \\    
        \hspace{.5cm} $(M/L)_\text{fil}$                        & 74.5      & 84.1 \\
        \hspace{.5cm} $(M/L)_\text{fil,bkg-sub}$                & 39.2      & 43.2 \\ 
        \hspace{.5cm} $\overline{(M/L)}_\text{cuts}$            & 80.6      & 88.0 \\ 
        \hspace{.5cm} $\overline{(ML)}_\text{cuts,bkg-sub}$     & 74.2      & 83.3 \\ \hline
    \textbf{Central density $n_0$ [10$^6$ cm$^{-3}$]}           &           &  \\
        \hspace{.5cm} $(n_0)_\text{avgProfile}$                 & 1.7       & 1.4 \\
        \hspace{.5cm} $\overline{(n_0)}_\text{cuts}$            & 1.8       & 1.5 \\ \hline
    \textbf{$R_\text{flat}$ [au]}                               &           &  \\
        \hspace{.5cm} $(R_\text{flat})_\text{avgProfile}$       & 2600      & 3100 \\
        \hspace{.5cm} $\overline{(R_\text{flat})}_\text{cuts}$  & 3300      & 3800 \\ \hline
    \textbf{$FWHM_\text{d}$ [au]}                               &           &  \\
        \hspace{.5cm} $(FWHM_d)_\text{avgProfile}$              & 6200      & 6900 \\
        \hspace{.5cm} $\overline{(FWHM_d)}_\text{cuts}$         & 6500      & 7200 \\ \hline
    \textbf{Exponent $p$}                                       &           &  \\
        \hspace{.5cm} $(p)_\text{avgProfile}$                   & 2.91      & 2.98 \\
        \hspace{.5cm} $\overline{(p)}_\text{cuts}$              & 3.40      & 3.51 \\ \hline
    \textbf{Kinetic temperature $T_\text{kin}$ [K]}\tablenotemark{a} & 8.7  &  9.6\\
    \textbf{Velocity dispersion $\sigma_{\text{v,NH}_3}$ [km s$^{-1}$]}\tablenotemark{a}    & 0.117          &  0.104 \\ \hline   
    \enddata
    \tablenotetext{a}{Averaged value along the spine of the filament.}
\end{deluxetable}

\section{Discussion}\label{sec:discussion}

\subsection{Similarities between spine profiles}\label{ssec:sims}

\begin{figure*}[th]
    \centering
    \includegraphics[width=0.98\textwidth]{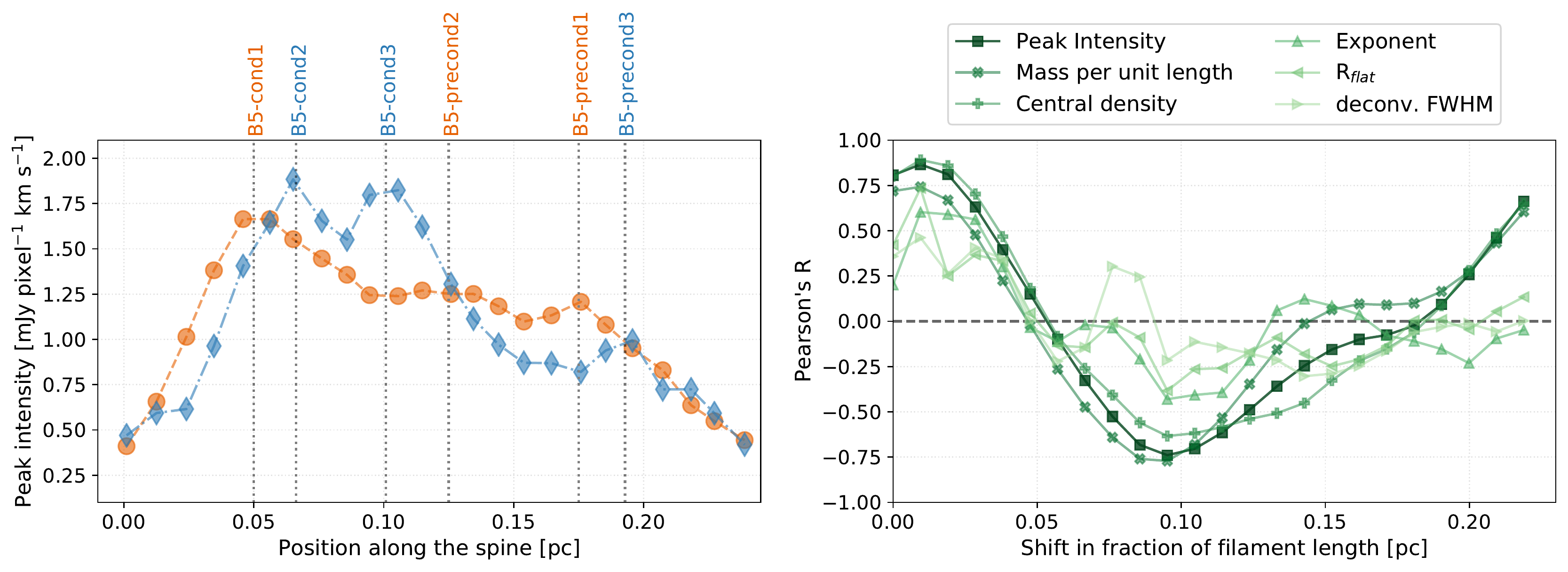}
    \caption{\emph{Left:} Same as Fig. \ref{fig:B5_fil_properties} Panel (a), but the spine profile of B5-fil2 is inverted and scaled. \emph{Right:} Correlation of different spine profiles between B5-fil1 and B5-fil2 as a function of relative shift along the spine. See Sec. \ref{ssec:sims} for details.}
    \label{fig:similarities}
\end{figure*}

We noticed that visually the spine profiles of B5-fil1 and B5-fil2 appear similar if one starts following them from the end at which their embedded condensations are located and scales them to the same length. To quantify this further, we have inverted the spine profiles of B5-fil2 shown in Fig.~\ref{fig:B5_fil_properties}, i.e. we follow the spine of this filament from North to South. We rescale the filament length by a factor of 0.767 to match it to the length of B5-fil1. We then interpolated the spine profiles of B5-fil2 and extract profile values at the same equidistant positions as for B5-fil1. An example of the resulting spine profiles of this procedure is shown for the peak intensity spine profile on the right side in Fig.~\ref{fig:similarities}. We then calculate the Pearson's R statistic between both filaments for each of the following properties: peak intensity, mass per unit length $M/L$, central density $n_0$, Plummer-exponent p, $R_\text{flat}$ and FWHM$_\text{d}$. We apply a shift to the spine profiles of B5-fil2 and calculate the Pearson's R statistics to find the best alignment between the filaments. The variation of the Pearson's R values as a function of shift is shown on the left side in Fig.~\ref{fig:similarities}. The variation of the correlation is similar for all parameters. Shifting the spine profiles of B5-fil2 by \unit[0.01]{pc} to the North of this filament results in a high correlation between all parameters. This shift aligns the position of B5-cond1 with B5-cond2.

\subsection{Comparison with \emph{Herschel} filaments}\label{ssec:widths}

The filaments we study here are embedded within the coherent core of Barnard 5. In this subsection we compare the properties derived for these filaments to the properties of the  `\emph{Herschel}' filaments.

The \emph{Herschel} Gould Belt Survey \citep[HGBS,][]{Andre2010} has mapped star-forming regions at five different wavelengths (\unit[70 -- 500]{\micron}), from which the HGBS team has derived H$_2$ column density maps at the native resolution of \unit[$~\sim$36]{\arcsec} and at a medium resolution of \unit[18.4]{\arcsec} following the procedure presented in \citet{Palmeirim2013}. In their recently published study, \citet{Arzoumanian2019} investigate eight of these star forming regions and identify in total $\sim$~600 individual filaments that have an aspect ratio $>3$. For this sample of filaments 
they derive distributions of filament properties (see their Tab. 3), such as the length of the filaments L~=~\unit[0.66 $\pm$ 0.46]{pc}, the mass per unit length $M/L$ = \unit[14 $\pm$ 18]{\msun pc$^{-1}$}, the central column density N$_{H_2}^0$~=~\unit[(7.0 $\pm$ 6.2)~$\times$~10$^{21}$]{H$_2$ cm$^{-2}$}, the plummer profile exponent $p$~=~2.2~$\pm$~0.3, and the deconvolved filament width FWHM$_d$~=~\unit[0.1 $\pm$ 0.05]{pc}.

We compare this to the results from fitting the averaged profiles of the B5 filaments (see Tab. \ref{tab:filProps}). The B5 filaments are short (L~=~\unit[0.28]{pc}), more super-critical ($M/L$~=~\unit[41.2]{\msun pc$^{-1}$}), steeper (p~=~3.0), and narrower (FWHM$_d$~=~\unit[0.03]{pc}) than the \emph{Herschel} filaments.

When using molecular lines to study filaments, properties such as the inferred filament widths seems to be dependent on the observed molecule. For 55 dense fibers identified in Orion's Integral Shape Filament, \citet{Hacar2018} find the distribution of filament width peaking at \unit[0.035]{pc} using N$_2$H$^+$. \citet{Panopoulou2014} on the other hand infer a broad distribution of filament width peaking at \unit[0.4]{pc} in the Taurus molecular cloud using $^{13}$CO. Studying the Orion A molecular cloud using C$^{18}$O, \citet{Suri2019} finds a varying filament width between 0.02 and \unit[0.3]{pc}, mostly in agreement with the dust continuum based study by \citet{Arzoumanian2019}. They attribute the large spread to the amount of substructure present within a filament.
In many of these cases however, the spatial resolution of the molecular line data used to determine the filament widths is often better by a factor of 2 -- 4 compared to the medium-resolution H$_2$ column density map derived from the \emph{Herschel} continuum observations.

The difference in the filament properties found for the B5 filaments could be due to  different angular resolution or the fact that ammonia is a high density tracer and as such picks out only the spine of the filaments.

\textit{Angular resolution}: We smooth the high-resolution NH$_3$(1,1) integrated intensity map to the medium-resolution of the \emph{Herschel} column density map, i.e. \unit[18.4]{\arcsec}. We use the filament spine determined using fil\_finder on the high-resolution NH$_3$(1,1) integrated intensity map and extract equidistant cuts perpendicular to the spines. We average the cuts and apply the same fitting procedure as described in Sec.~\ref{ssec:fitting}. The details of this processing as well as the results are given in Appendix~\ref{app:width}.
The averaged profiles of the B5 filaments become even steeper (p = 4.14), and while their width increases, they remain narrow (FWHM$_d$ = \unit[0.05]{pc}).

\textit{High-density tracer versus continuum}: We use the medium-resolution H$_2$ column density map derived for Perseus that is publicly available on the HGBS webpage\footnote{http://www.herschel.fr/cea/gouldbelt/en}. We apply the same processing as for the smoothed NH$_3$(1,1) integrated intensity map described above. The details of this processing are given in Appendix~\ref{app:width}.
The averaged profiles of the B5 filaments remain steep (p = 3.33), their widths increases marginally (FWHM$_d$ = \unit[0.06]{pc}). The central column density is higher (N$_{H_2}^0$ = \unit[1.3 $\times$ 10$^{22}$]{H$_2$ cm$^{-2}$}) compared to the \emph{Herschel} filaments. There is only very little difference between the fitting results of smoothed high-density tracer and the H$_2$ column density map, indicating that NH$_3$(1,1) is not filtering out the extended wings of filaments and hence is a good tracer of filamentary structures.

In summary, the physical properties (length, width, volume density, density profile) of the filamentary substructure present within the coherent core of B5 are significantly different from those measured in \emph{Herschel} filaments. This is not surprising, considering the higher density environment within which the B5 filamentary substructure has formed, and it suggests that environmental conditions play a crucial role in shaping the physical characteristics of filaments in general.

\subsection{Can the filaments be magnetically supported?}\label{ssec:bfield}

Both filaments are supercritical, but the turbulence present in the coherent core (sonic Mach number, $M_s = \sigma_\text{nt} / c_s$ =  0.5) is insufficient to support them against gravitational collapse. Hence they should be in a state of collapse -- unless magnetic fields are present, which could provide additional support against gravitational collapse \citep{Fiege2000,Seifried2015}.
This poses the question of how strong a magnetic field would need to be to stabilize the filaments. 

We present here a first attempt to estimate the magnetic field strength in B5, which has not yet been measured.
A toroidal field does not stabilize the filament against radial collapse \citep{Fiege2000}. A poloidal field that is perpendicular to the spine of the filament has also been found to not stabilize it \citep{Seifried2015}. This leaves a poloidal field that is parallel to the filament, which has indeed been found to stabilize a filament \citep{Fiege2000, Seifried2015}. We use this orientation to estimate the mininum magnetic field required to fully stabilize the filaments against collapse, which would ultimately prevent any fragmentation happening.

The critical mass per unit length for an isothermal cylinder is given in Eq.~\ref{eq:isoMline} \citep{Ostriker1964}. The cylinder can be further stabilized by turbulence and magnetic field. Taking the turbulence into account, \citet{Fiege2000} derive a modified critical mass per unit length (see their Eq. 12):

\begin{equation}\label{eq:MLcrit2}
    (M/L)_\text{crit,nt} = \frac{2\sigma_\text{v}^2}{G},
\end{equation}
where $\sigma_\text{v}$ is the velocity dispersion of the gas including thermal and non-thermal (nt) motions (see Eq.~\ref{eq:sigma_v}).

Taking the magnetic field $B_z$ parallel to the spine of the filament into account, \citet{Fiege2000} determine:
\begin{equation}\label{eq:MLcrit3}
    (M/L)_\text{crit,mag} = \frac{2\sigma_\text{v}^2 + \frac{B_z^2}{4 \pi \mu_\text{p} m_\text{H} n_0}}{G},
\end{equation}
\textbf{where $\mu_\text{p}=2.37$ is the mean molecular weight per free particle \citep{Kauffmann2008}, and $m_\text{H}$ is the hydrogen atom mass.}
Assuming that our filaments are gravitationally stable, i.e. the observationally measured mass per unit length $(M/L)_\text{obs}$ equals $(M/L)_\text{crit,mag}$, evaluating Eq.~\ref{eq:MLcrit3} for the magnetic field $B_z$, and using Eq.~\ref{eq:MLcrit2} yields
\begin{equation}\label{eq:Bfield}
    B_z = \sigma_v \sqrt{8 \pi \mu_p m_\text{H} n_0 \left( \frac{(M/L)_\text{obs}}{(M/L)_\text{crit,nt}} - 1\right)}.
\end{equation}

\begin{figure*}[th]
    \centering
    \includegraphics[width=0.98\textwidth]{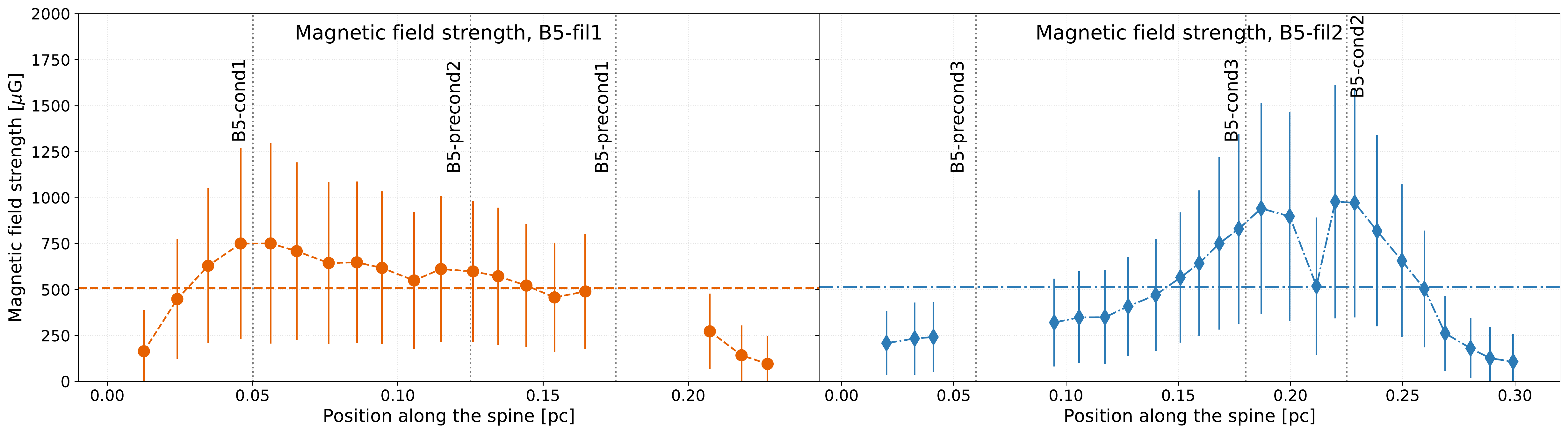}
    \caption{Local variation of the estimated magnetic field strength $B_z$ required to stabilize the filament for B5-fil1 (\emph{left}) and B5-fil2 (\emph{right}). The dashed and dash-dotted horizontal lines mark the average values of the magnetic field strength. The vertical gray dotted lines mark the location of the condensations as well as the pre-condensations. The position 0.0 starts at the South of the spine. The holes are due to in the kinetic temperature being undetermined at the respective positions.}
    \label{fig:bfield}
\end{figure*}

In Fig.~\ref{fig:bfield} we show the distribution of required magnetic field strength $B_z$ to support the filaments. 
The mean $B_z$ is \unit[510]{$\upmu$G} for B5-fil1 and \unit[515]{$\upmu$G} for B5-fil2. We notice an increase in required magnetic field strength towards the condensations up to values of \unit[$\sim$950]{$\upmu$G}. Since fragmentation seems to be already ongoing towards the condensations this is not unexpected. 
We exclude the tips of the filaments as they deviate from the cylinder assumption. We note that the holes in the profiles originate from the kinetic temperature being undetermined at the respective position.

Another region, also embedded in the Perseus molecular cloud and hence sharing the same parental molecular cloud as B5, is Barnard 1 (B1). The B1 clump hosts several pre- and protostellar cores at different evolutionary stages. Recently, \citet{Coude2019} determine the magnetic field strength of B1 to be \unit[120]{$\upmu$G} using JCMT polarization data. 
\citet{Chapman2011} report the plane-of-sky magnetic field strength in the nearby star-forming region Taurus to range from 10 to \unit[40]{$\upmu$G} using near-infrared polarization observations. \citet{Sugitani2011}, on the other hand determine a rough estimate of the magnetic field strength in the Serpens cloud to be a \unit[$\text{few}\times100$]{$\upmu$G}. 
\citet{Liu2019} report magnetic field strengths towards a low-mass starless core in the $\rho$ Ophiuchus cloud of \unit[103 - 213]{$\upmu$G} using three different methods and JCMT \eight dust polarization observations. Also towards Ophiuchius, \citet{Pattle2020} report magnetic field strengths ranging between \unit[72 -- 366]{$\upmu$G}.

Our inferred value of the magnetic field strength is an upper limit. It exceeds many of the measured values in other (low-mass) star-forming regions or clumps. So while it may still be possible that the filaments could be marginally supported by magnetic pressure, additional observations are required to quantify this.

\begin{figure*}[th]
    \centering
    \includegraphics[width=0.98\textwidth]{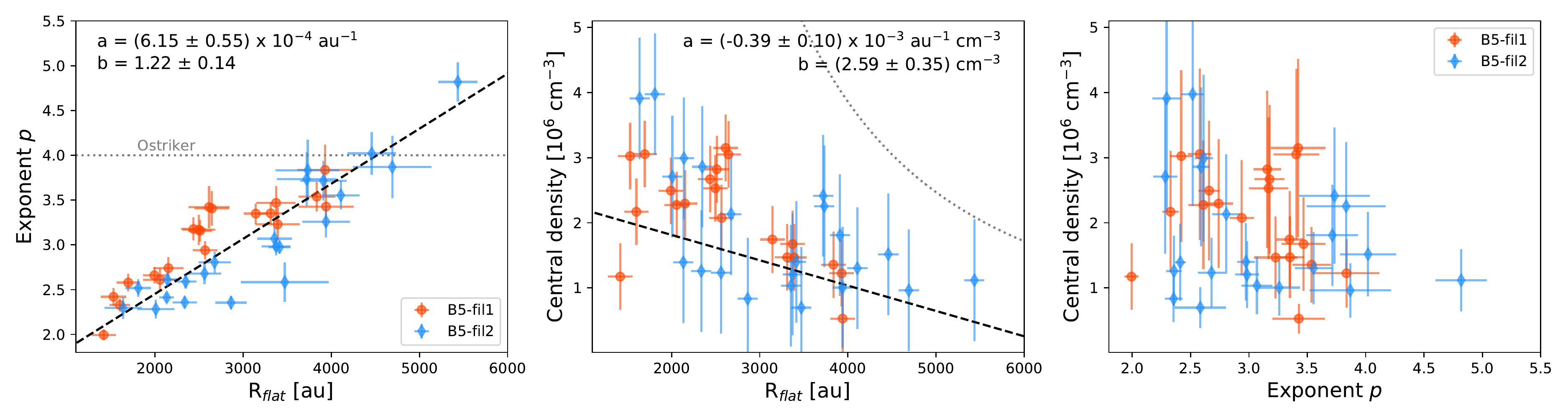}
    \caption{\emph{Left:} Correlation between exponent $p$ and flattening radius $R_\text{flat}$. \emph{Middle} Anti-correlation between central density $n_0$ and flattening radius $R_\text{flat}$. The dotted line marks the critical length  $\lambda_\text{crit} = 3.94 H$, with $H$ as given in Eq. \ref{eq:H} \citep{Ostriker1964}. \emph{Right:} The relation between central density $n_0$ and exponent $p$.}
    \label{fig:correlation}.
\end{figure*}

\subsection{Fragmentation}

The simplest case of cylindrical fragmentation is that of an isothermal, pressure-supported, infinitely-long filament. The gravitational fragmentation of such a system has a critical wavelength of $\lambda_\text{crit} = 3.94\,H$, where
\begin{equation}\label{eq:H}
    H^2 = \frac{2\,c_s^2}{\pi\, G\, \mu\, n_0},
\end{equation}
\citep{Hacar2011, Stodolkiewicz1963,Ostriker1964}.
Taking the average central density of the filaments of $n_0 =$ \unit[$10^6$]{$\text{cm}^{-3}$} and a gas temperature of \unit[9]{K}, yields a critical wavelength of $\lambda_\text{crit} =$ \unit[7400]{au} (\unit[0.04]{pc}). We determine the closest separation between two structures along the filament spines to be \unit[0.045]{pc}, between B5-cond2 and B5-cond3. The separation between B5-cond1 and B5-precond2 along the filament spine is \unit[0.05]{pc}. 

But this approach neglects the influence of turbulence and magnetic fields on the critical wavelength. \citet{Fiege2000a} investigate the effect of the magnetic field and its orientation on the critical wavelength. For a poloidal magnetic field parallel to the filament spine, they find that the scale for the separation of the fragments increases with increasing magnetic field strength. This could mean that some of our identified substructure within the filaments is spaced too closely. However, to further investigate the influence of the magnetic field on the separation of the fragments will require a measurement of the magnetic field orientation and the magnetic field strength.

\subsection{Filament Evolution}\label{ssec:evolution}

As mentioned in Sec.~\ref{ssec:width_exp}, we notice a relation between $R_\text{flat}$, $n_0$ and the exponent $p$ along the spines of these filaments. To investigate this further, we plot the relation between all three parameters with each other in Fig.~\ref{fig:correlation}. Especially clear is the correlation between the exponent $p$ and $R_\text{flat}$. The anti-correlation between the central density and $R_\text{flat}$ is less clear due to the high error associated with the central density, to which the uncertainty in the temperature determination contributes the most. The best-fit linear fits shown in Fig.~\ref{fig:correlation} take the associated uncertainties into account. The emerging trend is that dense filaments are narrower and filaments deviate from the fiducial isothermal filament model as they become denser and narrower.

\begin{figure}[tbp]
    \centering
    \includegraphics[width=0.45\textwidth]{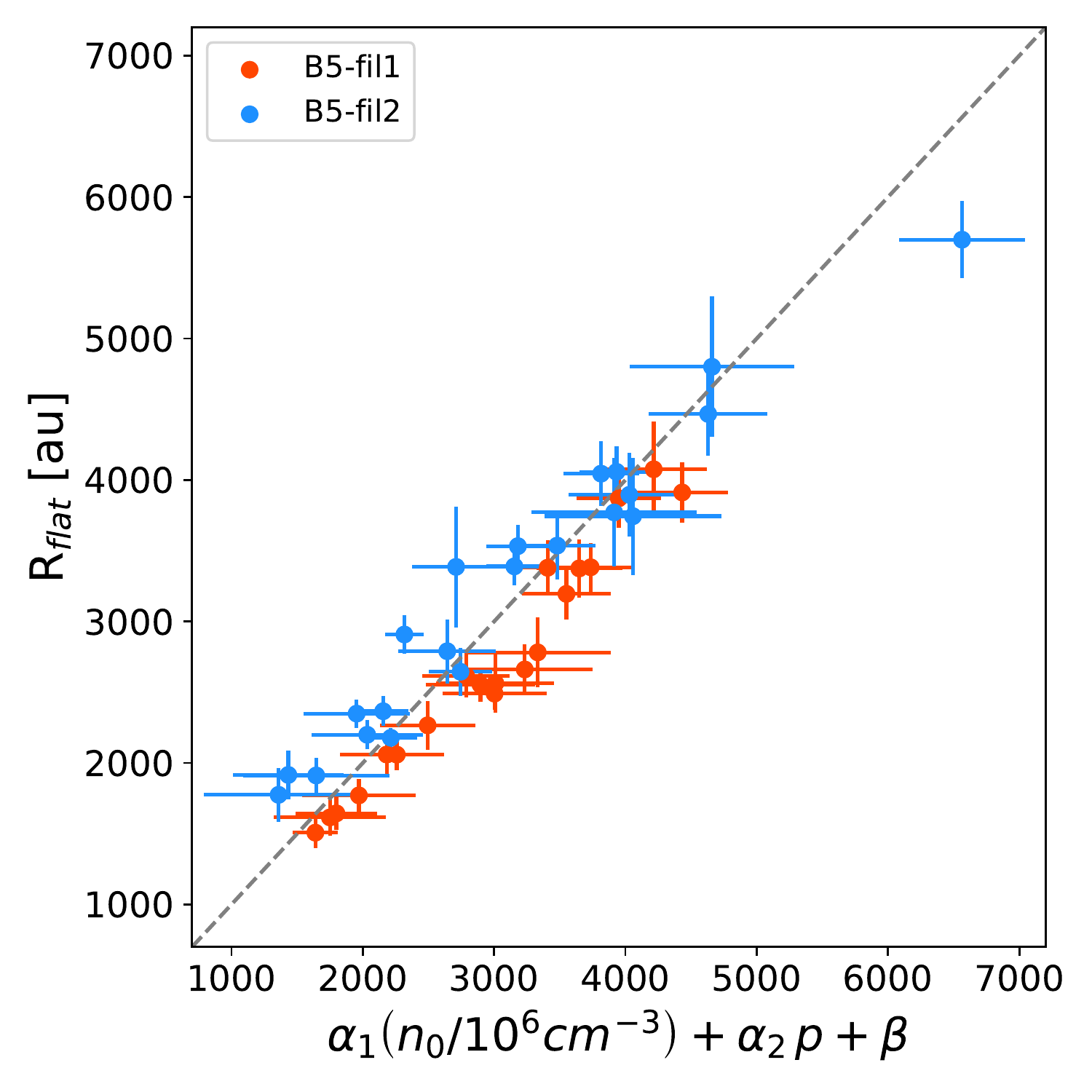}   
    \caption{Correlation between the flattening radius $R_\text{flat}$, central density $n_0$, and exponent $p$ along the filament spines.}
    \label{fig:hyperfit}
\end{figure}

To derive an empirical relation between all three parameters, we fit a plane to the three-dimensional parameter space,

\begin{equation}
    R_\text{flat} = \alpha_1 \, \frac{n_0}{10^6 \text{cm}^{-3}} + \alpha_2 \,  p + \beta,
\end{equation}
where $\alpha_1$ and $\alpha_2$ are scaling parameters and $\beta$ is the intercept of the plane.
Since all three parameters have a measurement error associated with them, we use the Hyper-fit package \citep{Robotham2015} to fit a plane in 3D. This package allows to fit linear models to multi-dimensional data with multivariate Gaussian uncertainties and provides access to a multitude of fitting algorithms. We employ the Nelder-Mead algorithm and obtain the following values: $\alpha_1$~=~-320~$\pm$~71, $\alpha_2$~=~1544~$\pm$~111, and $\beta$~=~-1153~$\pm$~360. Fig.~\ref{fig:hyperfit} shows the reprojected hyperplane.

This provides the first observational prescription of the evolution of the physical parameters of the filaments dependent on $R_\text{flat}$. Although this relation might not be universal, it shows the first clear evolution of a filament that is collapsing or fragmenting revealing the global trends.

For an isothermal filament in hydrostatic equilibrium (exponent $p = 4$), \citet{Arzoumanian2011} state that the flattening radius $R_\text{flat}$ corresponds to the thermal Jeans length $\lambda_\text{J}$, which is anti-correlated to the central column density $\Sigma_0$. They investigate three low-mass star forming filaments, Aquila, IC5146, and Polaris and find a lack of anti-correlation between the filament widths, which they determine as the deconvolved FWHM from the Gaussian fit to the profile, and the column density (see their Fig.~7). However, from fitting a Plummer-like function to the mean column density profile of the filaments they determine the exponent p to range between 1.5 and 2.5, i.e. less steep than the exponents of B5-fil1 and B5-fil2. They determine a flattening radius $R_\text{flat} = $ \unit[2000 -- 16000]{au}, i.e. only a small subsection has similar values as B5-fil1 and B5-fil2. 
Similarily, \citet{Suri2019} analyze C$^{18}$O observations of the high-mass star forming region Orion A. They determine the widths of the 625 individual, relatively short (\unit[$<$1.7]{pc}), filaments that they identify. Using the column density map derived from \emph{Herschel} data \citep{Stutz2015}, they find no (anti-)correlation between filament widths and column density (see their Fig.~11). On the other hand they do find a correlation between filament widths and number of shoulder in the radial intensity profiles detected, suggestive of (unresolved) substructures within the filaments. Therefore it is possible that their analysis is e.g. affected by optical depth effects or the peak column density could be underestimated due to the large \emph{Herschel} beam.

It could be possible that ammonia is tracing higher density material and hence we are able to see the expected inverse relation between $R_\text{flat}$ and the central density $n_0$. A similar investigation for other filaments would be needed to confirm the relationship observationally. In addition numerical simulations are required to investigate if and under which circumstances these correlations appear.

\section{Summary}\label{sec:summary}

In this paper, we analyze previously published combined VLA and GBT ammonia data together with JCMT continuum data of Barnard 5. Embedded in the coherent region we find two clumps, one quiescent and the other one containing two filamentary structures. Embedded in the filaments are three condensations and along the spines of the filaments we find signs of additional clumpy structures. We characterize the filament properties in detail by fitting both a Plummer function and a Gaussian to equidistant cuts extracted perpendicular to the filament spines.

\begin{itemize}
    \item Both filaments are narrow and dense. Their deconvolved FWHM range between \unit[6200 -- 7000]{au}, i.e. 2$\times$ the flattening radius of the Plummer function, and their average central density is on the order of \unit[10$^6$]{cm$^{-3}$}. Their aspect ratios are 17:1 and 10:1. 
    \item The line-of-sight velocity is \unit[$\sim$10.2]{km s$^{-1}$} and the velocity dispersion along the filament spines is \unit[0.1]{km s$^{-1}$}, but increases towards the location of the condensations.
    \item Both filaments are super-critical, exhibiting mass per unit length values of \unit[$\sim$ 80]{\msun pc$^{-1}$}. Locally this value increases up to \unit[150]{\msun pc$^{-1}$}.
    \item We estimate the required magnetic field strength to stabilize the filaments, ultimately stopping further fragmentation, to be on the order of \unit[$\sim$500]{$\upmu$G}. Since we see signs of ongoing fragmentation, we conclude that this magnetic field strength is an upper limit.
    \item We fit the radial profiles perpendicular to the filament spines and see a variation in power-law exponent and width along the filament. Their maxima are coincident with the peak positions of the condensations, which could be related to the filament evolution. 
    \item We find a strong correlation between the Plummer exponent and the flattening radius. We also find an anti-correlation between the central density and this flattening radius, suggestive of contraction. The measurements of these three parameters (central density, Plummer exponent and flattening radius) fall in a plane and we derive their empirical relation. Numerical simulations are needed to see if and under what circumstances these correlations are being seen.

\end{itemize}

%
\acknowledgments
We thank the anonymous referee for their helpful comments that improved the paper. 
A.S., J.E.P., P.C., M.J.M and D.S.C. acknowledge the support of the Max Planck Society.
G.A.F acknowledges financial support from the State Agency for Research of the Spanish MCIU through the AYA2017-84390-C2-1-R grant (co-funded by FEDER) and through the “Center of Excellence Severo Ochoa” award for the Instituto de Astrofísica de Andalucia (SEV-2017-0709).
The National Radio Astronomy Observatory and the Green Bank Observatory are facilities of the National Science Foundation operated under cooperative agreement by Associated Universities, Inc.
This research has made use of data from the Herschel Gould Belt survey (HGBS) project (http://gouldbelt-herschel.cea.fr). The HGBS is a Herschel Key Programme jointly carried out by SPIRE Specialist Astronomy Group 3 (SAG 3), scientists of several institutes in the PACS Consortium (CEA Saclay, INAF-IFSI Rome and INAF-Arcetri, KU Leuven, MPIA Heidelberg), and scientists of the Herschel Science Center (HSC).

\facilities{VLA, GBT, JCMT}

\software{aplpy \citep[v.2.0.3;][]{Robitaille2012},
          astrodendro \citep[v0.2.0;][]{Rosolowsky2008},
          astropy \citep[v4.0.2;][]{Astropy2013, AStropy2018},
          fil-finder \citep[v1.7;][]{Koch2015},
          hyper-fit \citep{Robotham2015},
          lmfit \citep[v1.0.1;][]{Newville2016},
          numpy \citep[v.1.19.2;][]{VanDerWalt2011},
          pyspeckit \citep[v0.1.23;][]{Ginsburg2011},
          radfil \citep[v1.1.2;][]{Zucker2018a},
          spectral-cube \citep[v0.5.0;][]{Robitaille2016}
          }



\def\url#1{}
\def\dodoi#1{}
\def\doarXiv#1{}
\def\doeprint#1{}

%

\vspace{1cm}
\bibliographystyle{aasjournal}
\bibliography{b5.bib}

\appendix

\section{Line fitting} \label{app:line_fitting}

We simultaneously fit the NH$_3$(1,1) and (2,2) lines using the \verb+cold-ammonia+ model \citep{Friesen2017} in \verb+pySpecKit+ \citep{Ginsburg2011}. This model assumes that both transition have the same excitation temperature, $T_{\rm ex}$ and that only the (1,1) and (2,2) levels are populated, i.e. the model does not present hyperfine anomalies \citep{Stutzki1984}. The best fit is obtained by minimizing the $\chi^2$, which also provides an uncertainty for each parameter. Input parameter guesses are based on (a) velocity centroid of the line for $v_{lsr}$, (b) intensity-weighted second moment of the velocity around the centroid for the velocity dispersion, $\sigma_v$, (c) $\log_{10}(N/\textrm{cm}^{-2}$) = 14.5, (d) $T_K$ = \unit[12]{K}, and (e) $T_{ex}$ = \unit[3]{K}, based on the temperature of the cosmic-ray background radiation (\unit[2.73]{K}). We include all pixels in the fit where the NH$_3$(1,1) line has a S/N ratio $\ge$ 5. 
We account for the channel response by applying the following correction to the velocity dispersion: 
\begin{equation}
    \sigma = \sqrt{\sigma_\text{v,fit}^2 - \frac{dv_\text{chan}}{2\sqrt{2ln2}}},
\end{equation}
where $\sigma_\text{v,fit}$ is the velocity dispersion from the fit and $dv_\text{chan}$ is the channel width.

We perform additional masking to remove poor fit results on a pixel-by-pixel basis for each parameter map. The line-of-sight velocity and velocity dispersion can be reliably obtained with a good fit of the NH$_3$(1,1) line, therefore we flag only those pixels with an associated uncertainty of \unit[$>$0.02]{km s$^{-1}$} in $v_{lsr}$ or $\sigma_v$.
Determination of the excitation and kinetic temperatures, $T_{ex}$ and $T_K$, as well as the ammonia column density $N_{\text{NH}_3}$, requires a good fit for both, the NH$_3$(1,1) and the NH$_3$(2,2) transitions. An inaccurate kinetic temperature yields an erroneous excitation temperature. Hence we require a S/N ratio $\ge$ 12 for the NH$_3$(1,1) line and a S/N ratio $\ge$ 3 for the NH$_3$(2,2) line, as well as the associated uncertainties of the temperature fits $\sigma_\text{fit,T}$ \unit[$<$2]{K}. For the uncertainty of the ammonia column density, we require $\sigma_{\text{fit,}log_{10}(N/\textrm{cm}^{-2})} < 1.0$. In addition to that, an accurate determination of the ammonia column density depends on good determination of both temperatures, $T_{ex}$ and $T_K$ \citep{Ho1983,Friesen2009}. Hence, if a pixel has been flagged in the temperature maps, it will also be flagged in the column density map.

\begin{figure*}[t]
    \centering
    \includegraphics[width=0.99\textwidth]{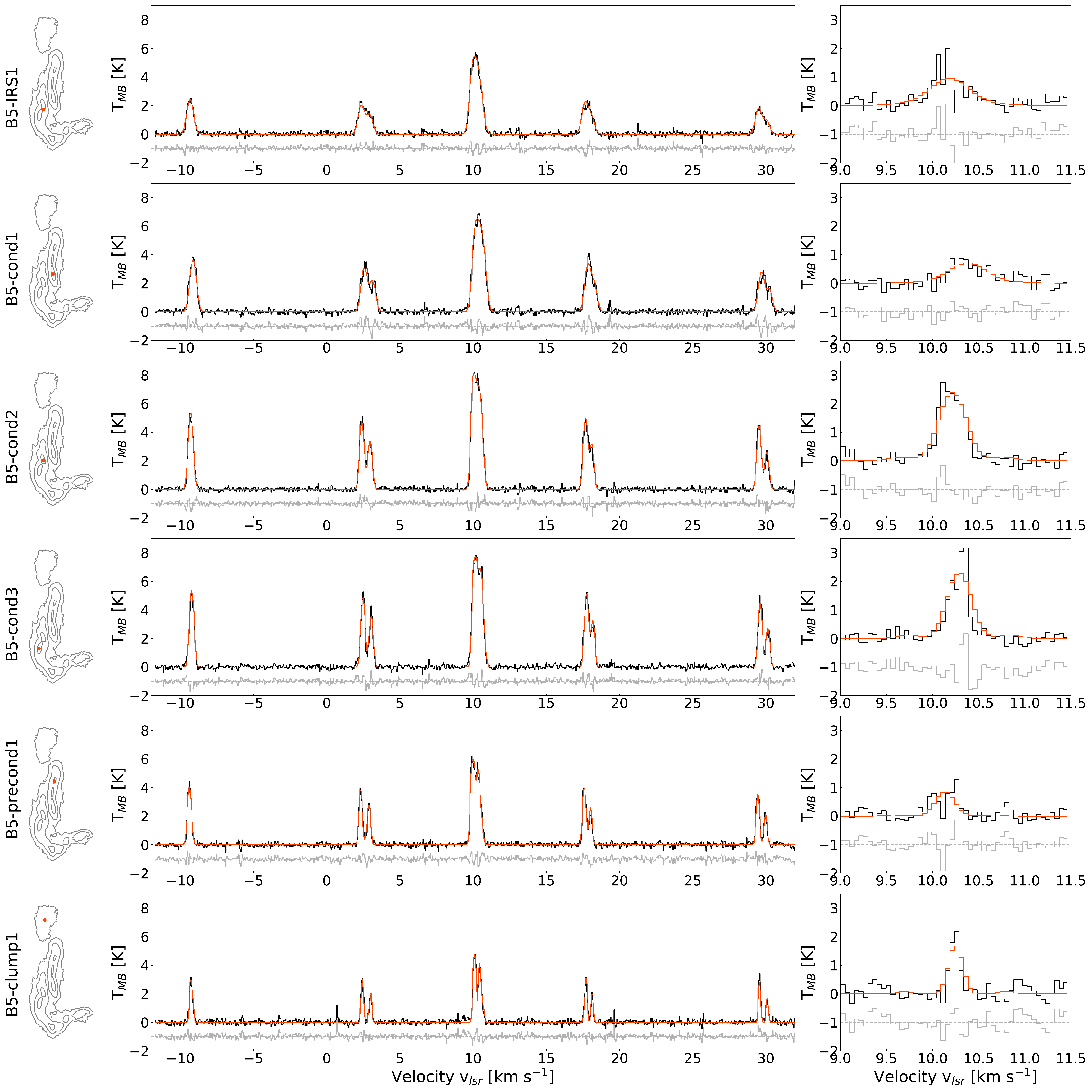}
    \caption{\emph{Top to bottom:} Sample spectra of ammonia towards six different positions in B5. \emph{Left column:} Contour map of B5 where the orange dot marks the location towards which the spectra have been extracted. \emph{Middle column:} Beam-averaged spectrum of NH$_3$(1,1). The observed spectrum is plotted in black, the best-fit model is shown in orange. The hyperfine splitting is clearly visible in each spectrum. \emph{Right column:} Beam-averaged spectrum of NH$_3$(2,2).} 
    \label{fig:nh3_spectra_fit}
\end{figure*}

Fig.\ref{fig:nh3_spectra_fit} shows the NH$_3$(1,1) and NH$_3$(2,2) beam-averaged spectra of six regions in the map, corresponding to the YSO B5-IRS1, the three condensations, B5-precond1, and B5-clump1, . Their respective locations are indicated by the orange dot in the contour map. The spectra clearly show the hyperfine splitting in the NH$_3$(1,1) transition thanks to the narrow velocity dispersions.
The final maps of ammonia column density $N_{\text{NH}_3}$, kinetic temperature $T_K$, center velocity $v_{lsr}$, and velocity dispersion $\sigma_v$ are shown in Fig.\ref{fig:B5_nh3_fit_propertyMaps}.


\section{Conversion factors}\label{app:convFactors}

\subsection{Ammonia-to-mass conversion by scaling}

We measure the flux density in B5-cond1 in the JCMT \four map, which is bright and without a protostar, to estimate its mass. 
To determine the total mass of the filaments, we use the JCMT \four map to scale the NH$_3$(1,1) integrated emission map, which is more sensitive to the filamentary structure. In the first step, we calculate the flux-to-mass conversion factor $\xi$ based on the conditions in B5-cond1 using:
\begin{equation}
\xi = \mathcal{G} \frac{d^2}{\kappa_\nu B_\nu(T_d)},    
\end{equation}
where $\mathcal{G}$ is the gas-to-dust ratio \citep[$\mathcal{G} = 100.$;][]{Hildebrand1983}, $d$ is the distance to the object \citep[$d = (302 \pm 21)\,$pc;][]{Zucker2018}, $\kappa_\nu$ is the dust absorption coefficient at the frequency $\nu$, and $B_\nu(T_d)$ is the Planck function evaluated at the dust temperature $T_d$.

To determine the dust temperature, we fit a Gaussian to the distribution of kinetic temperatures within the filaments (see Sec. \ref{app:line_fitting}), and obtain a mean temperature of \unit[$T_\text{k} = (9 \pm 1)$]{K}. This assumes that the dust temperature is coupled to the gas temperature. 
We also assume that the dust is optically thin, covered in thin to thick ice mantles, and coagulated at \unit[($10^5$~--~$10^7$)]{cm$^{-3}$}. We interpolate the corresponding tabulated dust opacities provided by \citet{Ossenkopf1994} to a wavelength of \four and then determine a mean dust absorption coefficient $\kappa_{450}$=\unit[(6.4~$\pm$~0.8)]{cm$^2$ g$^{-1}$}. With that, we derive a SCUBA flux-to-mass conversion factor of $\xi$=\unit[(0.5~$\pm$~0.2)]{\msun Jy$^{-1}$}.

We determine the background emission for both the JCMT \four map and the integrated NH$_3$(1,1) map as the lowest-value pixel within the contour of B5-cond1. We note that this will yield very conservative mass estimates. For B5-cond1 we measure a background corrected flux density of \unit[0.99]{Jy} in the JCMT \four map. For the same condensation, we measure a background corrected flux density of \unit[0.35]{Jy} in the integrated NH$_3$(1,1) map.  This yields an ammonia-to-mass conversion factor of \unit[(1.5~$\pm$~0.7)]{\msun Jy$^{-1}$}.

\subsection{Ammonia-to-mass conversion from column density}\label{sssec:mass_coldens}

We calculate the mass from the ammonia column density map on a pixel-by-pixel basis using
\begin{equation}
    M = \mu_{H_2}\, m_H\, A_\text{pixel}\, \frac{N_{NH_3}}{X_{NH_3}},    
\end{equation}
where $\mu_{H_2}$ is the molecular weight per hydrogen molecule \citep[$\mu_{H_2}=2.8$; ][]{Kauffmann2008}, $m_H$ is the atomic hydrogen mass, $A_\text{pixel}$ is the size of the pixel (assuming a distance of \unit[d=302]{pc}), $N_{NH_3}$ is the ammonia column density from the fit (see SubSec. \ref{app:line_fitting}), and $X_{NH_3}$ is the abundance of ammonia with respect to $H_2$. For a sample of clouds with similar conditions, \citet{Friesen2017} determined the abundance of ammonia with respect to $H_2$ to be $X_{NH_3} = 10^{-8.5}$ on average.

\subsection{Comparison}
The resulting mass maps for the filaments agree within a factor of two with each other. The mass map derived from the column density is a factor of 2 higher compared to the mass map derived by scaling the continuum. The residual between both maps is smooth and does not show any strong gradients.
All mass-dependent calculations in this work were performed on the basis of the mass-map derived with the dust-scaling method.

\begin{figure*}[ht]
    \centering
    \includegraphics[width=0.98\textwidth]{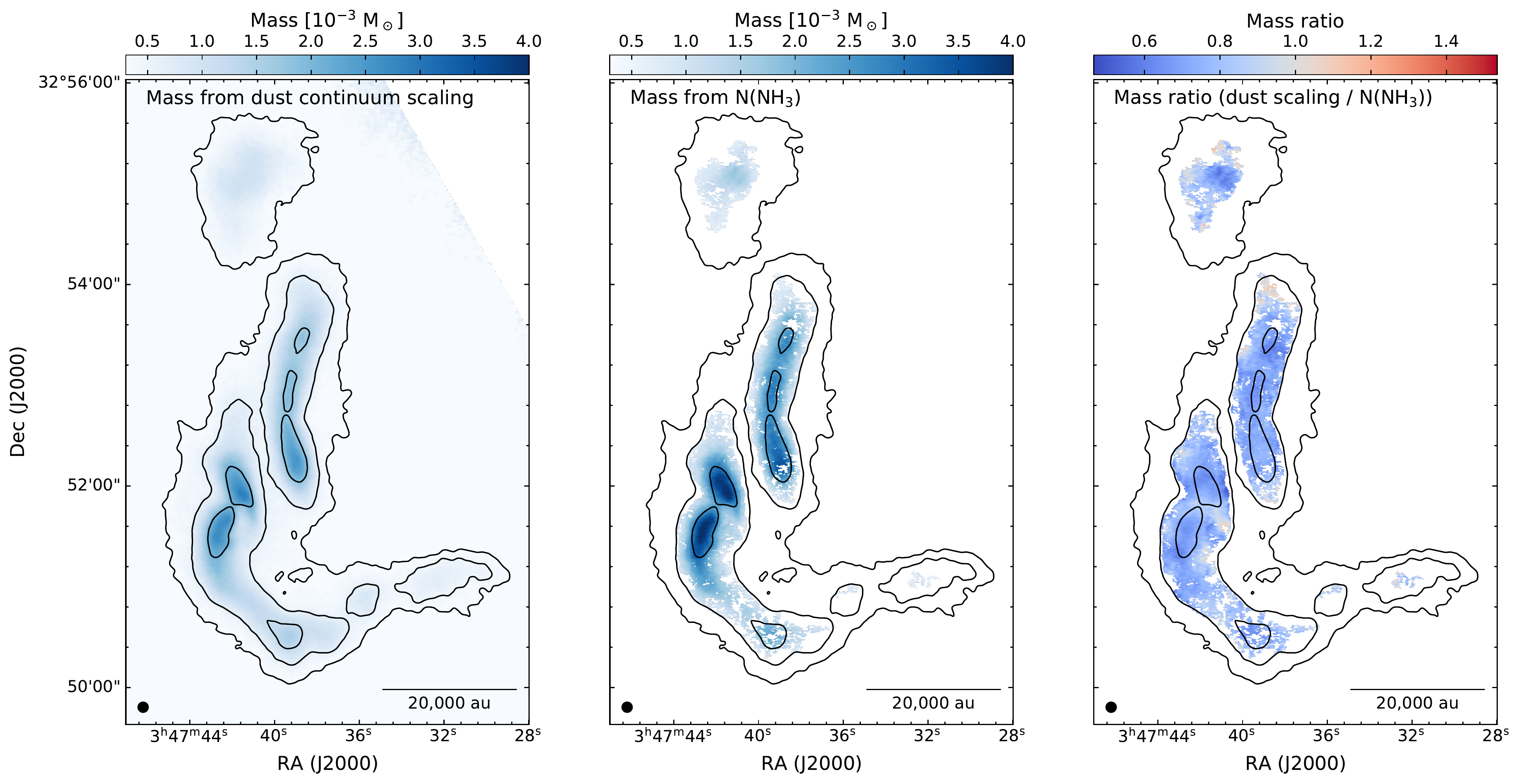}
    \caption{\emph{Left:} Mass map derived from scaling the VLA+GBT NH$_3$ integrated intensity map based on the mass scaling factor derived from the JCMT \four dust continuum map.
    \emph{Middle:} Mass map derived from the NH$_3$ column density maps obtained from the line fitting (see App.~\ref{app:line_fitting}). \emph{Right:} Ratio map of the left and middle map. The scalebar is shown in the bottom right corner of each map.} 
    \label{fig:mass_comparison}
\end{figure*}

\section{Comparison of filament width at \unit[18.2]{\arcsec} resolution}\label{app:width}

B5 was observed with \emph{Herschel} PACS and SPIRE as part of the \emph{Herschel} Gould Belt Survey \citep[HGBS;][]{Andre2010}. H$_2$ column density maps at two different spatial resolutions (\unit[36]{\arcsec} and \unit[18.2]{\arcsec}) are available on the HGBS webpage. The original resolution H$_2$ column density map was derived from SED fitting of the \unit[160 -- 500]{\micron} maps on a pixel-to-pixel basis \citep[for details see e.g.][]{Pezzuto2020,Koenyves2015,Pezzuto2012}. The \unit[18.2]{\arcsec} resolution H$_2$ column density map was determined using a multi-scale decomposition technique as described in \citet{Palmeirim2013}.

\begin{figure*}[ht]
    \centering
    \includegraphics[width=0.9\textwidth]{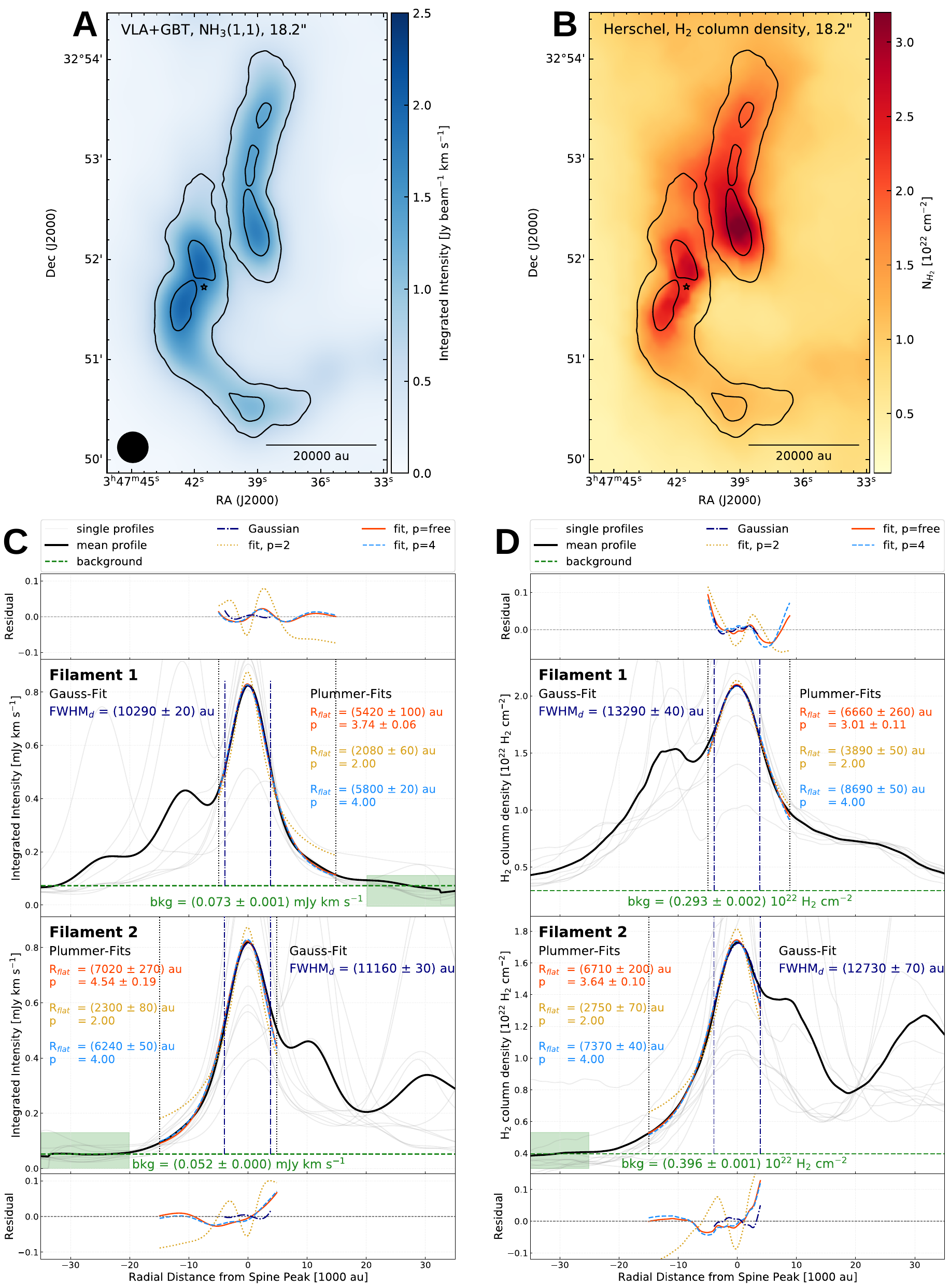}
    \caption{\emph{Panel A}: NH$_3$(1,1) integrated intensity map convolved to a spatial resolution of \unit[18.2]{\arcsec}. \emph{Panel B}: H$_2$ column density map at a resolution of \unit[18.2]{\arcsec} derived from \emph{Herschel} observations \citep{Palmeirim2013}. \emph{Panel C}: Same as Fig.~\ref{fig:filamentProfiles}, but for a lower spatial resolution. \emph{Panel D}: Same as Fig.~\ref{fig:filamentProfiles}, but for the H$_2$ column density map at \unit[18.2]{\arcsec} resolution.} 
    \label{fig:widths_comparison}
\end{figure*}

We convolve our VLA+GBT NH$_3$(1,1) integrated intensity map to the same resolution as the medium-resolution H$_2$ column density map. Both maps are shown in the top row of Fig.~\ref{fig:widths_comparison}. We use the same method as described in Sec.~\ref{ssec:fitting} to extract equidistant cuts along the filament spines. We calculate the mean of the cuts and first fit a constant to determine the background emission level, followed by fitting a Gaussian profile to the innermost part. For the profiles of the H$_2$ column density map, we adjust the range for the the background emission fitting, since the emission is more extended compared to the NH$_3$ integrated intensity map. The fitting ranges for the Gaussian and Plummer fits are adjusted slightly to exclude contamination from nearby structures. The fit ranges are indicated with vertical lines in panels C and D in Fig.~\ref{fig:widths_comparison}. The fit results are summarized in Tab.~\ref{tab:widths_comparison} and discussed in Sec.~\ref{ssec:widths}.

\begin{deluxetable}{c|cl|llll|c}[]
    \tablecaption{Results of fitting the Plummer and Gaussian functions to the averaged profiles for three data sets. \label{tab:widths_comparison}}
    \tablehead{Source & Fit-function & Data set &{Parameters} & & & & Evaluation}
    \startdata
    \textbf{B5-fil1}    & \textbf{Plummer}\tablenotemark{a} & & \textbf{p} & \textbf{R$_\text{flat}$} [au] & \textbf{A}\tablenotemark{b}  & \textbf{bkg}\tablenotemark{b,c} & \textbf{AIC} \\ 
        & & NH$_3$, \unit[6]{\arcsec}      & 2.0\tablenotemark{d} & 1220 $\pm$ \phn40 & 1.184 $\pm$ 0.019 & 0.060 $\pm$ 0.001 & -1150 \\
        & & NH$_3$, \unit[18.2]{\arcsec}   & 2.0\tablenotemark{d} & 2080 $\pm$ \phn60 & 0.826 $\pm$ 0.013 & 0.073 $\pm$ 0.001 & -1132 \\       
        & & \emph{Herschel}, \unit[18.2]{\arcsec} & 2.0\tablenotemark{d} & 3890 $\pm$ \phn50 & 1.858 $\pm$ 0.008 & 0.293 $\pm$ 0.002 & \phn-828 \\
        \cline{3-8}
        & & NH$_3$, \unit[6]{\arcsec}      & 2.91 $\pm$ 0.06      & 2590 $\pm$ \phn80 & 1.090 $\pm$ 0.007 & 0.060 $\pm$ 0.001 & -1462 \\
        & & NH$_3$, \unit[18.2]{\arcsec}   & 3.74 $\pm$ 0.06      & 5420 $\pm$ 100    & 0.767 $\pm$ 0.002 & 0.073 $\pm$ 0.001 & -1741 \\       
        & & \emph{Herschel}, \unit[18.2]{\arcsec} & 3.01 $\pm$ 0.11      & 6660 $\pm$ 260    & 1.818 $\pm$ 0.005 & 0.293 $\pm$ 0.002 & \phn-980 \\
        \cline{3-8}        
        & & NH$_3$, \unit[6]{\arcsec}      & 4.0\tablenotemark{d} & 3790 $\pm$ \phn40 & 1.052 $\pm$ 0.008 & 0.060 $\pm$ 0.001 & -1349 \\
        & & NH$_3$, \unit[18.2]{\arcsec}   & 4.0\tablenotemark{d} & 5800 $\pm$ \phn20 & 0.764 $\pm$ 0.002 & 0.073 $\pm$ 0.001 & -1729 \\       
        & & \emph{Herschel}, \unit[18.2]{\arcsec} & 4.0\tablenotemark{d} & 8690 $\pm$ \phn50 & 1.802 $\pm$ 0.004 & 0.293 $\pm$ 0.002 & \phn-948 \\
    \cline{2-8}
        & \textbf{Gaussian}\tablenotemark{a} & & \textbf{A$_\text{G}$} & \textbf{$\bm{\sigma}_\text{G}$} [au] & \textbf{$\bm{\mu}_\text{G}$} [au] & \textbf{bkg}\tablenotemark{b,c} & \textbf{AIC}\\           
        & & NH$_3$, \unit[6]{\arcsec}      & \phn6449 $\pm$ 23 & 2530 $\pm$ \phn10 & \phn-54 $\pm$ 9    & 0.060 $\pm$ 0.001 & -640\\
        & & NH$_3$, \unit[18.2]{\arcsec}   & \phn6946 $\pm$ 16 & 3700 $\pm$ \phn10 & \phn$\,\,$40 $\pm$ 6 & 0.073 $\pm$ 0.001 & -805\\
        & & \emph{Herschel}, \unit[18.2]{\arcsec} & 23153 $\pm$ 74    & 5140 $\pm$ \phn20 & -127 $\pm$ 8       & 0.293 $\pm$ 0.002 & -718\\
    \hline
    \textbf{B5-fil2}  & \textbf{Plummer}\tablenotemark{a} & & \textbf{p} & \textbf{R$_\text{flat}$} [au] & \textbf{A}\tablenotemark{b} & \textbf{bkg}\tablenotemark{b,c} & \textbf{AIC} \\
        & & NH$_3$, \unit[6]{\arcsec}      & 2.0\tablenotemark{d} & 1470 $\pm$ \phn40 & 1.134 $\pm$ 0.017 & 0.0494 $\pm$ 0.0004 & -1161 \\
        & & NH$_3$, \unit[18.2]{\arcsec}   & 2.0\tablenotemark{d} & 2300 $\pm$ \phn80 & 0.841 $\pm$ 0.014 & 0.0517 $\pm$ 0.0003 & -1061 \\       
        & & \emph{Herschel}, \unit[18.2]{\arcsec} & 2.0\tablenotemark{d} & 2750 $\pm$ \phn70 & 1.443 $\pm$ 0.018 & 0.3964 $\pm$ 0.0009 & \phn-892 \\
        \cline{3-8}       %
        & & NH$_3$, \unit[6]{\arcsec}      & 2.98 $\pm$ 0.05 & 3140 $\pm$ \phn80 & 1.049 $\pm$ 0.006 & 0.0494 $\pm$ 0.0004 & -1537 \\
        & & NH$_3$, \unit[18.2]{\arcsec}   & 4.85 $\pm$ 0.29 & 7020 $\pm$ 270    & 0.774 $\pm$ 0.005 & 0.0517 $\pm$ 0.0003 & -1482 \\       
        & & Herschel, \unit[18.2]{\arcsec} & 3.64 $\pm$ 0.10 & 6710 $\pm$ 200    & 1.359 $\pm$ 0.005 & 0.3964 $\pm$ 0.0009 & -1295 \\
        \cline{3-8}           %
        & & NH$_3$, \unit[6]{\arcsec}      & 4.0\tablenotemark{d} & 4440 $\pm$ \phn40 & 1.016 $\pm$ 0.006 & 0.0494 $\pm$ 0.0004 & -1412 \\
        & & NH$_3$, \unit[18.2]{\arcsec}   & 4.0\tablenotemark{d} & 6240 $\pm$ \phn50 & 0.785 $\pm$ 0.004 & 0.0517 $\pm$ 0.0003 & -1473 \\       
        & & \emph{Herschel}, \unit[18.2]{\arcsec} & 4.0\tablenotemark{d} & 7370 $\pm$ \phn40 & 1.351 $\pm$ 0.005 & 0.3964 $\pm$ 0.0009 & -1287 \\
     \cline{2-8}
        & \textbf{Gaussian}\tablenotemark{a} & & \textbf{A$_\text{G}$} & \textbf{$\bm{\sigma}_\text{G}$} [au] & \textbf{$\bm{\mu}_\text{G}$} [au] & \textbf{bkg}\tablenotemark{b,c} & \textbf{AIC}\\           
        & & NH$_3$, \unit[6]{\arcsec}      & \phn7063 $\pm$ 29 & 2830 $\pm$ \phn20 & $\,$-39 $\pm$ 10    & 0.0494 $\pm$ 0.0004 & -627\\
        & & NH$_3$, \unit[18.2]{\arcsec}   & \phn7904 $\pm$ 20 & 4130 $\pm$ \phn10 & 136 $\pm$ \phn7 & 0.0517 $\pm$ 0.0003 & -826\\
        & & \emph{Herschel}, \unit[18.2]{\arcsec} & 16227 $\pm$ 94    & 4880 $\pm$ \phn30 & 201 $\pm$ 15    & 0.3964 $\pm$ 0.0009 & -662\\
    \enddata
    \tablenotetext{a}{The fitting ranges where identical and kept fixed.}
    \tablenotemark{b}{Units for NH$_3$ map are mJy km s$^{-1}$; units for column density map derived from \emph{Herschel} continuum maps are 10$^{22}$H$_2$ cm$^{-2}$.}
    \tablenotetext{c}{The background (bkg) has been determined by first fitting a constant and was fixed in the subsequent Plummer/Gaussian fit.}
    \tablenotetext{d}{The parameter has been fixed.}

\end{deluxetable}

\section{Assessing degeneracy in fitting high-contrast Plummer profiles}\label{app:fitQuality}

We examine the reliability of the filament properties using our fitting method outlined in Sec.~\ref{ssec:fitting}. To this end we generate a suite of synthetic profiles, with given $R_\text{flat}$, exponent $p$, and peak intensity $A$ to sample the parameter space uniformly. We set the constant background level to the mean of the background level of the fitted profiles in this work (\unit[0.06]{mJy}). We add Gaussian noise with a $\sigma$ of \unit[0.02]{mJy}, derived from the integrated intensity map of NH$_3$(1,1) and convolve the synthetic profile with the \unit[6]{\arcsec}-Gaussian beam. We uniformly sample the parameter space of $R_\text{flat}$ between \unit[2000]{au} and \unit[10000]{au}, and the exponent $p$ between 2.0 and 6.0. 
\citet{Arzoumanian2019} define the contrast $C^0$ of a profile as 
\begin{equation}
    C^0 = \frac{I_\text{peak} - I_\text{bkg}}{I_\text{bkg}},
\end{equation}
where $I_\text{peak}$ and $I_\text{bkg}$ are the peak and background intensity of the profile, respectively.
For the correlation plot in Fig.~\ref{fig:correlation}, we exclude all perpendicular profiles that are at the tips of the filaments. For the remaining filaments, the contrast ranges between 8 and 60. Values larger than 2.0 are commonly considered high-contrast profiles. To test the same contrast regime, we set the peak intensity $A$ between \unit[0.54]{mJy} and \unit[3.66]{mJy}. In total we sample 4,000 unique parameter sets and fit each of them for 20 different noise seeds.

Input parameter guesses are based on the fit of a Gaussian to the innermost part of each profile, where (1) the Gaussian amplitude is used as the initial guess of the Plummer $A$ parameter, and (2) the Gaussian $\sigma$ divided by 2 is used a an initial guess for $R_\text{flat}$. For the exponent we set the initial guess to 3 (the mean of the parameter space). We employ the same fitting range as for the profile fitting performed in the paper, i.e. we fit the profile between \unit[-6000 and 15000]{au}. Hence we cut off one shoulder of the profile and include the major part of the shoulder on the other side. We convert the fitting parameter $A$ to central density using Eq.~\ref{eq:convCentralDensity}. 

We perform additional masking to remove poor fit results on a profile-by-profile basis. We remove all profiles with an associated uncertainty \unit[$>$1500]{au} in flattening radius $R_\text{flat}$, $>1.2$ in the exponent $p$, and \unit[$>$ 3.4 $\times$ 10$^{6}$][cm$^{-3}$] in central density $n_0$. These are three times the maximum of the uncertainty of each parameter in the fit of the observational profiles.

\begin{figure*}[ht]
    \centering
    \includegraphics[width=0.98\textwidth]{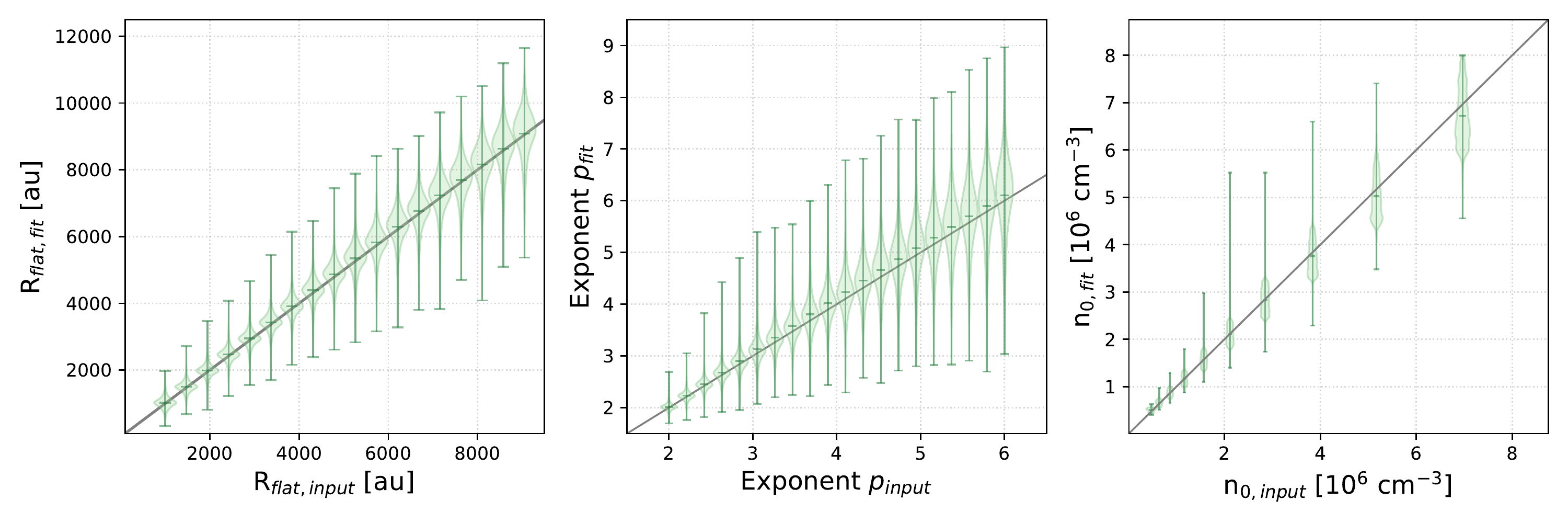}
    \caption{Relation between fitted parameters and input parameters of synthetic Plummer profiles. \emph{Left:} R$_\text{flat}$. \emph{Middle:}  Exponent $p$. \emph{Right:} Central density $n_0$. The solid grey line mark the 1:1 relation.} 
    \label{fig:synFit}
\end{figure*}

In Figure~\ref{fig:synFit} we show the relation between each fitted parameter and the corresponding input parameter. For all parameters, the fitted values recover the input values within the uncertainties. The uncertainties increase towards larger values of $R_\text{flat}$ and steeper profiles. This has mainly to do with the profiles becoming wider and the fitting range cutting off before the shoulder is reached. However, the mean of all fit recovers the input values.

\end{document}